%% file: main.tex
\documentclass{aifrontiers}
\usepackage{aifrontiers}
\usepackage{booktabs}
\usepackage{array}
\usepackage{colortbl}
\usepackage{caption}
\usepackage{subcaption}
\usepackage{multirow}
\usepackage{multicol}
\usepackage{float}
\usepackage{enumitem}
\usepackage{xspace}
\usepackage{soul}
\usepackage{listings}
\usepackage[export]{adjustbox}
\usepackage{algorithm}
\usepackage{algorithmic}
\usepackage{tikz}
\usetikzlibrary{positioning,arrows.meta,fit,calc,backgrounds}
\newif\ifusehtmlfig
\usehtmlfigtrue

\input{preamble}

\newcommand{\DRA}{\textsc{DRA}\xspace}
\newcommand{\NP}{\mathrm{NP}}

\newcommand{\revert}{\ominus}
\newcommand{\Units}{\mathcal{U}}
\newcommand{\Oracle}{\mathcal{O}}
\newcommand{\Eres}{E_{\mathrm{dbg}}}
\newcommand{\Efix}{E_{\mathrm{fix}}}
\newcommand{\role}[1]{\textsc{#1}}

\newif\ifshowplaceholders
\showplaceholdersfalse
\definecolor{fillbg}{RGB}{255, 240, 120}
\definecolor{actionbg}{RGB}{255, 200, 150}

\begin{document}

\title{How Much Do Coding Agents Over-Edit, and Why? An Empirical Study of Unneeded Edits in Repository-Level Agent Patches}

\shorttitle{Unneeded Edits in Coding-Agent Patches}

\author{
  {\bfseries Aakash Goel\thanks{Equal contribution.}, Benjamin Steenhoek\footnotemark[1], Yu Hu} \\[0.45em]
    {\normalsize Microsoft}\\[0.45em]
    {\normalsize \texttt{\{aakashgoel, bensteenhoek, yuhu\}@microsoft.com}}
}
\date{\today}
\renewcommand{\thefootnote}{\fnsymbol{footnote}}

\renewcommand{\weblink}{}
\renewcommand{\foundrylink}{}
\renewcommand{\hflink}{}
\renewcommand{\ghlink}{}

\input{content}

\end{document}

%% file: preamble.tex
\usepackage[T1]{fontenc}    %
\usepackage[utf8]{inputenc} %

\usepackage[capitalize,noabbrev]{cleveref}
\crefname{algorithm}{Algorithm}{Algorithms}
\Crefname{algorithm}{Algorithm}{Algorithms}
\newcommand{\tabref}[1]{\cref{#1}}
\newcommand{\figref}[1]{\cref{#1}}
\newcommand{\secref}[1]{\cref{#1}}
\newcommand{\needed}{\textsc{needed}}
\newcommand{\unneeded}{\textsc{unneeded}}

\newcommand{\tablestyle}[2]{\setlength{\tabcolsep}{#1}\renewcommand{\arraystretch}{#2}\centering\footnotesize}

\newcolumntype{x}[1]{>{\centering\arraybackslash}p{#1pt}}
\newcolumntype{y}[1]{>{\raggedright\arraybackslash}p{#1pt}}

\definecolor{baselinecolor}{gray}{0.93}
\newcommand{\baseline}[1]{\cellcolor{baselinecolor}{#1}}

\newif\ifreviewflags
\reviewflagsfalse
\DeclareRobustCommand{\staleflag}{\ifreviewflags\raisebox{0.5ex}{\tikz[baseline]{\draw[red!70!black,line width=0.3pt](0,0)--(0,1.05ex);\fill[red!80!black](0,1.05ex)--(0.75ex,0.8ex)--(0,0.55ex)--cycle;}}\fi}

\newcommand{\reviewlegend}{\ifreviewflags\noindent\textit{Review version: \staleflag{} marks content not updated for this result set.}\par\fi}

%% file: content.tex
\newcommand{\finding}[2]{\par\smallskip\noindent\fcolorbox{black!35}{black!4}{\parbox{\dimexpr\linewidth-2\fboxsep-2\fboxrule}{\small\textbf{Finding #1.} #2}}\par\smallskip}
\newcommand{\Ddev}{\Delta_{\mathrm{dev}}}

\begin{abstract}
Agent patches are larger than developers', which is widely read as over-editing. But patch size does not say whether the extra edits are needed, what they are, or where they come from. We introduce \textbf{Delta Residue Attribution} (\textbf{\DRA}), which replays agent trajectories, traces every block (unit) of the final patch to the step that wrote it, reverts each unit, and reruns the task tests to decide whether the unit is needed. We apply \DRA{} to 7{,}660 resolved GitHub Copilot CLI trajectories from 18 frontier models on SWE-bench Verified (6{,}728) and DeepSWE (932), and grade developer patches using the same method. Agent patches do change more lines than developers'; on SWE-bench Verified the majority of extra edits add tests (61\%) and documentation (13\%) while on DeepSWE, most extra edits are code (58\%). On both benchmarks, agents' code is on average at least as necessary as developers' (mean \(\NP_{\mathrm{code}}\) 0.887 vs.\ 0.864 on SWE-bench Verified, where 11 of 18 models leave less unneeded code than developers, and 0.797 vs.\ 0.785 on DeepSWE). On average, 11\% of an agent patch's code units on SWE-bench Verified are still unneeded, and 89\% of those change behavior. Tests the agent adds detect most injected bugs (68\%, against 77\% for the task tests); 5\% of all injected bugs are caught by the agent's new or strengthened tests but missed by the task tests. Debug steps write almost none of the unneeded code. Code necessity differs substantially by model; it is unrelated to resolve rate on SWE-bench Verified and negatively correlated with it on DeepSWE (Spearman \(-0.59\), \(p=0.01\)). On SWE-bench Verified, pruning with \DRA{} using only the tests the agent can see (its own tests and the repository's) removes the majority of unneeded code but also some needed code (67\% recall, 46\% precision), and 89\% of pruned patches still resolve the task; asking the agent to audit its own diff is more precise but less effective (36\% recall, 66\% precision). We recommend that benchmarks report code necessity next to resolve rate and that over-editing be addressed through model training and harness-level pruning.
\end{abstract}

\maketitle

\section{Introduction}
\label{sec:intro}

Language models are now routinely asked to edit existing code rather than write it from scratch. Passing tests is necessary but not sufficient: a patch that fixes a three-line bug by rewriting fifty lines is harder to review and more likely to carry unintended changes. In a maintainer study, roughly half of the SWE-bench-passing patches would not have been merged~\citep{metr2026merge}, and many plausible patches behave differently from the developer fix on inputs the tests do not cover~\citep{wang2026patchdiff}. We call the tendency to change more than necessary \emph{over-editing}. On minimal function-level edits, models change far more than the task requires~\citep{zhu2026overediting}. At the repository level, successful agent patches contain about twice as many changes as developers' patches, and minimality prompts barely help~\citep{luo2026recap}. Larger agent patches are therefore widely read as over-editing.

Preservation prompts~\citep{zhu2026overediting,chai2026pdb} and edit-aware training~\citep{zhu2026overediting,yang2026paft,ke2026prepair} reduce function-level over-editing. Patch size does not tell whether the extra edits are needed, what kind of edits they are, or when in the trajectory they were written; the answers to these questions determine what a fix should look like. If the extra edits are debugging leftovers, a harness can clean them up. If they are tests, they may be welcome. If they are written into the fix, the model has to change.

We study over-editing in repository-level patches at the level of individual blocks and the trajectory steps that wrote them. Our research questions are: \textbf{RQ1:} How much do agents over-edit compared with developers? \textbf{RQ2:} What are the unneeded edits? \textbf{RQ3:} Where in the trajectory do they come from? \textbf{RQ4:} Which models over-edit more, and does over-editing track resolve rate? \textbf{RQ5:} Can unneeded code be removed automatically?

To answer these questions, we build \textbf{\DRA}. \DRA{} replays a resolved trajectory, splits the final patch into units (contiguous blocks of changed lines that can be reverted on their own), and traces each unit to the step that wrote it. It decides whether a unit is needed by reverting it and rerunning the task tests, in the tradition of delta debugging~\citep{zeller2002dd}. The share of needed units is the \emph{necessity precision} \(\NP\), and \(\NP_{\mathrm{code}}\) is that share over code units. Because benchmark tests miss some needed behavior~\citep{wang2026patchdiff}, we also check unneeded code with behavioral probes (\secref{sec:oracle}) and grade developer patches the same way. We apply \DRA{} to 7{,}660 resolved coding-agent trajectories of 18 models on SWE-bench Verified and DeepSWE.

\reviewlegend

\paragraph{Findings.} Agent patches do change more lines than developers', but on SWE-bench Verified the extra lines are mostly tests (61\%) and documentation (13\%), while on DeepSWE most extra edits are code (58\%; RQ2). On code, on SWE-bench Verified 11 of 18 models leave less unneeded code than developers (mean \(\NP_{\mathrm{code}}\) 0.887 vs.\ 0.864), for four no significant difference is detected (which does not establish equivalence), and three models leave more (RQ1, \secref{sec:rq1}). On DeepSWE, agents are on par with developers (0.797 vs.\ 0.785), but no model is significantly above the developer baseline. On average, 11\% of an agent patch's code units are still unneeded, and 89\% of those change behavior (RQ2). Debug steps (exploring, reproducing, or probing) write almost none of the code residue (154 of 3{,}987 units); fix steps write 83\% of it (RQ3). Code necessity differs by model and does not correlate with resolve rate on SWE-bench Verified (Spearman \(-0.10\)), but on DeepSWE it is anti-correlated (\(-0.59\), \(p=0.01\); RQ4). On SWE-bench Verified, after generation, self-audit is precise (65.7\%) but finds only 35.5\% of unneeded code, pruning with the agent's own tests and the repository's regression tests finds 67.3\% at lower precision (46.3\%) on the 2{,}495 eligible patches, and stopping early barely changes \(\NP_{\mathrm{code}}\) (0.700 to 0.710; RQ5). Our contributions are a block-level empirical study of over-editing with developer patches graded the same way, the \DRA{} method (\secref{sec:method}), and implications for evaluation and harness design (\secref{sec:discussion}). Appendix~\ref{app:hypotheses} lists our pre-registered hypotheses and our deviations from them.

\section{Related Work}
\label{sec:related}

\paragraph{Measuring over-editing.}
\citet{zhu2026overediting} and PDB~\citep{chai2026pdb} measure unneeded edits against a known minimal fix, mostly on single functions, CanItEdit~\citep{cassano2024canitedit} measures how much of a function edit the tests cover, and edit-aware training reduces it~\citep{yang2026paft,ke2026prepair}. At the repository level, RECAP~\citep{luo2026recap} compares patch sizes. Coverage metrics conflate ``executed by tests'' with ``required by the task''. \DRA{} measures necessity by counterfactual removal and needs no unique reference patch.

\paragraph{Why unneeded changes are costly.}
Change size predicts defects~\citep{nagappan2005churn,kamei2013jit}, and tangled changes distort defect data~\citep{herzig2013tangled}. Reviewers favor small, focused changes~\citep{rigby2013convergent,bacchelli2013expectations}, and larger changes take longer to review~\citep{sadowski2018google}. Agent pull requests often need reviewer involvement or later rework~\citep{peralta2026agenticpr,pham2026agenticvshuman}. Plausible patches often differ from the developer fix on untested inputs~\citep{wang2026patchdiff}. LLM refactorings that pass existing tests can still change behavior~\citep{dristi2026difffuzz}. Both findings motivate our behavioral checks.

\paragraph{Agent harnesses, trajectories, and delta debugging.}
Work on agent harnesses and trajectories~\citep{yang2024sweagent,wang2025openhands,minisweagent,shu2026traceprobe,ceka2026traceability} does not attribute individual patch blocks to the steps that wrote them. \DRA{} builds on delta debugging~\citep{zeller2002dd} and inherits the overfitting caveat of test-based repair~\citep{smith2015cure}, so we calibrate against developer patches. Appendix~\ref{app:related} gives the extended discussion.

\section{Tracing Edits to Trajectory Steps: \DRA}
\label{sec:method}

\begin{figure}[t]
\centering
\includegraphics[width=0.9\columnwidth]{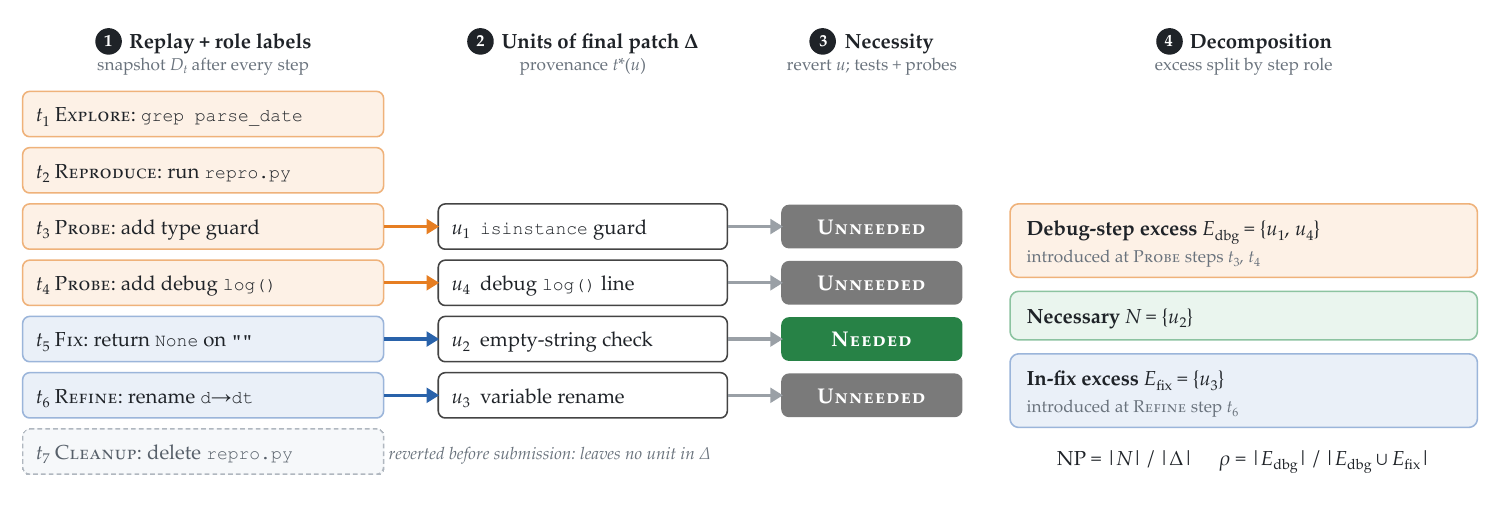}
\caption{Overview of \DRA. A resolved trajectory is replayed, each unit of the final patch is traced to the step that wrote it, the step is labeled with a role, and each unit is reverted to test whether it is needed. Unneeded units are attributed to the role of the step that wrote them. The example is the \texttt{parse\_date()} fix of Appendix~\ref{app:example}.}
\label{fig:pipeline}
\end{figure}

\DRA{} has four steps: replay, unit provenance, step-role labeling, and necessity grading. \figref{fig:pipeline} shows the pipeline on a small example (Appendix~\ref{app:example}), and \cref{alg:dra} gives pseudocode.

\begin{algorithm}[t]
\caption{Delta Residue Attribution (\DRA)}
\label{alg:dra}
\begin{algorithmic}[1]
\REQUIRE base files \(S_0\), resolved trajectory \(\tau=((a_t,o_t))_{t=1}^T\), submitted patch \(\Delta\), oracle \(\Oracle\)
\STATE replay the edit and create calls of \(\tau\) on \(S_0\), recording line-level blame after every step
\STATE assert the replayed final state equals \(\Delta\)
\FOR{each unit \(u \in \Units(\Delta)\)}
  \STATE \(t^\star(u) \leftarrow\) latest step that wrote or deleted a line of \(u\)
  \STATE \(r(t^\star(u)) \leftarrow \textsc{Rules}(a_{t^\star}) \text{ or } \textsc{Judge}(a_{t^\star}, o_{t^\star}, \text{context})\)
  \STATE \(\mathrm{class}(u) \in \{\text{code}, \text{test}, \text{docs}, \text{nonsemantic}, \text{config}\}\)
  \STATE \(n(u) \leftarrow 1 - \Oracle(S_T \revert u)\)
\ENDFOR
\RETURN \(\NP_{\mathrm{test}}(\Delta)\), \(\NP_{\mathrm{code}}(\Delta)\), \(\rho\), and the per-unit table
\end{algorithmic}
\end{algorithm}

\subsection{Definitions}
\label{sec:prelim}

An agent produces a trajectory of actions and observations; executing the first \(t\) actions of a task yields the working-tree state \(S_t\), and the submitted patch is \(\Delta = \mathrm{diff}(S_0,S_T)\). We decompose \(\Delta\) into units \(\Units(\Delta)=\{u_1,\dots,u_n\}\) that can each be reverted on their own, and write \(S_T \revert u\) for the state with unit \(u\) reverted. An oracle \(\Oracle\) accepts a state (\(\Oracle(S)=1\)) if it passes some check. The task-test oracle \(\Oracle_{\mathrm{test}}\) runs the SWE-bench FAIL\_TO\_PASS and PASS\_TO\_PASS tests~\citep{jimenez2024swebench} (on DeepSWE, the task's own tests). We analyze only resolved patches, so \(\Oracle(S_T)=1\). A unit is \emph{needed} for \(\Oracle\) if reverting it causes rejection (\(n_{\Oracle}(u)=1\), written \needed), and \emph{unneeded} otherwise (\(n_{\Oracle}(u)=0\), written \unneeded). The \emph{necessity precision} \(\NP_{\Oracle}\) of a patch is its needed units divided by all its units:

\noindent\begin{minipage}{0.46\columnwidth}
\begin{equation}
n_{\Oracle}(u) = 1 - \Oracle(S_T \revert u),
\label{eq:necessity}
\end{equation}
\end{minipage}\hfill
\begin{minipage}{0.52\columnwidth}
\begin{equation}
\NP_{\Oracle}(\Delta) = \frac{1}{|\Units(\Delta)|}\sum_{u} n_{\Oracle}(u).
\label{eq:np}
\end{equation}
\end{minipage}

\noindent \(\NP=1\) exactly when \(\Delta\) is 1-minimal~\citep{zeller2002dd}, meaning no single unit can be removed without the oracle rejecting the patch; larger groups of units may still be removable together. A \emph{unit} is an atomic piece of an edit. \emph{Unneeded code} is an unneeded unit in a code file; units in test files are graded separately (\secref{sec:units}). \(\NP\) inherits the oracle's blind spots: a unit that the tests do not need may still serve the developer's intent~\citep{smith2015cure}. We therefore treat the share of unneeded units as an upper bound on true excess, and we compare agents with developer patches graded the same way (Appendix~\ref{app:minimality} separates three notions of minimality).

\subsection{Replay, Units, and Provenance}
\label{sec:replay}\label{sec:units}\label{sec:provenance}

Agent trajectories record every tool call but not intermediate repository states. We replay the structured file-editing calls on the task's base files with line-level blame after each step, and we exclude trajectories whose replayed final state does not match the submitted patch exactly (96.7\% of SWE-bench Verified trajectories replay). A unit is a contiguous block of changed lines in one file (blocks separated by at most one unchanged line are merged), and a new or deleted file is one unit. Each unit is a \emph{code} unit or a non-code unit (test, documentation, configuration, or comment and formatting edits). SWE-bench grading resets test files, so test changes don't change the oracle's result. We therefore report test edits separately from code as \emph{ungraded}, and call ungraded test edits and unneeded units together the patch's \emph{extra edits}. We use \(\NP_{\mathrm{code}}\) as the primary measure (Appendix~\ref{app:replay} gives the rules). The introduction step \(t^\star(u)\) of a unit is the latest step that wrote or deleted any of its lines, so a unit first written at a debug step and later revised at a fix step is attributed to the fix.

\subsection{Step-Role Labeling}
\label{sec:roles}

Each step that introduces a unit gets a role. We group roles into three: \role{debug} (explore, reproduce, and probe steps), \role{fix} (fix and refine steps), and \role{cleanup} (steps that revert earlier changes). Two deterministic rules run first, and an LLM judge (GPT-6.1 Sol) labels the rest based on the issue, the condensed history, the action, and the agent's narration (Appendix~\ref{app:roles} gives the six fine-grained roles and the rubric).

\subsection{Oracle: Grading Necessity}
\label{sec:oracle}

For each unit we revert its lines in the task container and rerun the task tests. We count reversions that cannot be applied (they leave the base state or do not parse) as needed. Task tests do not exercise every input path, so we also check unneeded code units with behavioral probes modeled on PatchDiff~\citep{wang2026patchdiff}. We report necessity under \(\Oracle_{\mathrm{test}}\) and use the probes to validate its unneeded verdicts; the strengthened oracle \(\Oracle^{+}\), which also rescues units whose reversion matches the developer patch, uses the same probe labels. For each code unit that \(\Oracle_{\mathrm{test}}\) marks unneeded, an LLM (GPT-6.1 Sol) sees the issue and the diff of the reversion, without the task tests or the developer patch, and writes several differential calls: probe-script inputs that call the changed code and print labeled outputs. We run them under the agent patch, the reverted state, and the developer patch. A unit is needed under \(\Oracle^{+}\) if its reversion fails the task tests or makes the same behavioral change as the developer patch. Appendix~\ref{app:oracle} reports the breakdown. Units are reverted one at a time (\secref{sec:limitations}).

\subsection{Step Attribution and Developer Calibration}
\label{sec:residue}\label{sec:calibration}

The unneeded units form the residue \(\Eres \cup \Efix\): \(\Eres\) holds the units whose introducing step is a debug step, and \(\Efix\) holds the rest. The debug share is
\begin{equation}
\rho = |\Eres| \,/\, |\Eres \cup \Efix|,
\label{eq:rho}
\end{equation}
and \(\rho_{\mathrm{code}}\) restricts it to code residue. A debug-step edit that is needed is not residue, because the experiment became the fix.

Developer patches also contain code the tests do not need, so we grade SWE-bench Verified developer patches (code files only) the same way and report, for each model, the gap between the means
\begin{equation}
\Ddev = \overline{\NP_{\mathrm{code}}}(\text{agent}) - \overline{\NP_{\mathrm{code}}}(\text{developer})
\end{equation}
over instances where both have code units, with a 95\% instance-bootstrap CI (\(B=2000\); all CIs in this paper are 95\% bootstrap intervals over instances unless noted). If \(\Ddev<0\), the agent leaves more unneeded code than the developer.

\section{Study Setup}
\label{sec:experiments}

\subsection{Data}
\label{sec:data}

We evaluate 18 models from four providers on SWE-bench Verified~\citep{jimenez2024swebench,chowdhury2024swebenchverified} (500 Python coding tasks) and DeepSWE~\citep{huang2026deepswe} (113 long-horizon coding tasks in Go, TypeScript, Python, Rust, and JavaScript), using the GitHub Copilot CLI~\citep{githubcopilotcli} coding agent. SWE-bench Verified resolve rates range from 64.0\% to 89.4\%. The resolve rate of a model is the share of its attempted tasks whose final patch passes the task tests. For each model we use all of its resolved SWE-bench Verified instances whose trajectory replays exactly, and all of its resolved DeepSWE tasks. Model-level DeepSWE statistics use the 17 models with at least 10 graded tasks. The study covers 6{,}728 SWE-bench Verified patches (30{,}265 units; 482 distinct instances, 114 shared by all 18 models) and 932 DeepSWE patches (26{,}209 units). We ran developer baselines on SWE-bench Verified (470 analyzed instances, 1{,}317 code units) and on DeepSWE (103 graded tasks; four developer patches fail the task tests and are dropped). Other harnesses and SWE-bench Pro are reported in Appendix~\ref{app:ext}. All units are graded in the official evaluation images under \(\Oracle_{\mathrm{test}}\); behavioral probes validate the unneeded verdicts (\secref{sec:oracle}). Role and probe judges are GPT-6.1 Sol. A second judge (Claude Opus 5.5) relabeled 10{,}000 randomly sampled steps and agreed with the primary judge on 84.4\% of role labels (\(\kappa=0.69\)) and 89.1\% of debug vs.\ non-debug labels (\(\kappa=0.55\); Appendix~\ref{app:roles}).

\subsection{Research Questions and Analyses}
\label{sec:hypotheses}\label{sec:analyses}

For each research question, we pre-registered the outcome we expected (Appendix~\ref{app:hypotheses}), with thresholds and deviations from the original experiment plan. A two-model pilot showed that most extra edits are test edits, so we implemented separate analyses for code and tests, made code units the primary population, and added the developer calibration.

\section{Results}
\label{sec:results}

Appendix~\ref{app:models} lists per-model results and Appendix~\ref{app:ext} holds the secondary analyses referenced below.

\subsection{RQ1: How Much Do Agents Over-Edit?}
\label{sec:rq1}

By size, agent patches look bloated. On SWE-bench Verified the median model changes 2.33 times as many lines as the developer patch, in line with \citet{luo2026recap}. On code units alone the ratio is 0.86, and the per-task median is 1.00. Outside test files, agents still change more lines than developers (median ratio 1.09). By necessity, agents' mean \(\NP_{\mathrm{code}}\) is 0.887, against 0.864 for developer patches on the same instances. So agents produce more non-test changes, but most of the code they write is needed. Developer patches serve as a baseline for the oracle: \(\Oracle_{\mathrm{test}}\) alone calls 33\% of developer code units unneeded, and the strengthened oracle \(\Oracle^{+}\) calls 3\% unneeded (Appendix~\ref{app:oracle}).

\begin{figure}[t]
\centering\includegraphics[width=\linewidth]{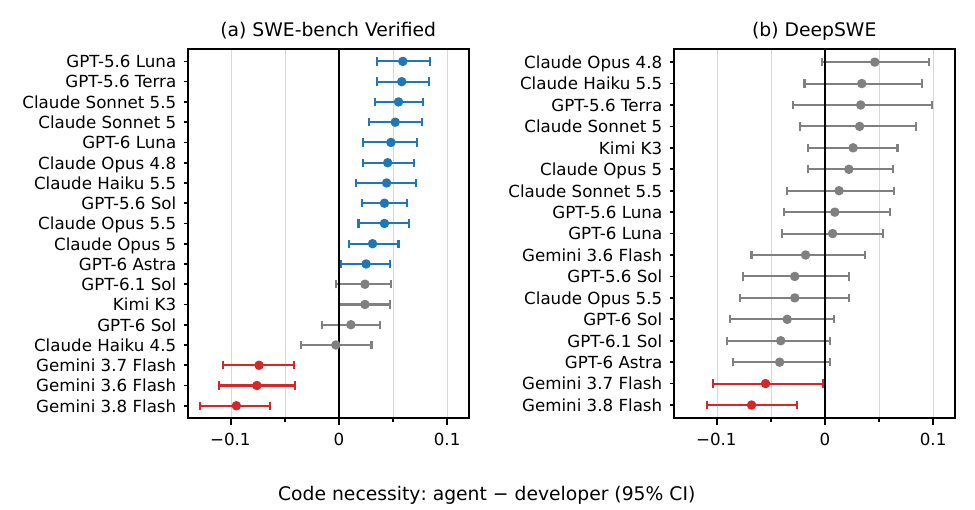}
\caption{Agent minus developer \(\NP_{\mathrm{code}}\) (\(\Ddev\)) per model under \(\Oracle_{\mathrm{test}}\), with 95\% bootstrap CIs. (a)~SWE-bench Verified instances with developer patches (6{,}728 agent patches, 470 developer patches). (b)~DeepSWE tasks with developer patches (914 agent patches from 17 models, 103 developer patches). Each panel is sorted by \(\Ddev\). Red: the CI excludes 0 below; blue: above.}
\label{fig:deltagold}
\end{figure}

\figref{fig:deltagold} shows the gap \(\Ddev\) on both benchmarks. On SWE-bench Verified (\figref{fig:deltagold}a), eleven models leave significantly less unneeded code than developers (GPT-6 Astra only marginally, with a CI lower bound of 0.002), four show no significant difference (Kimi K3, Claude Haiku~4.5, GPT-6 Sol, GPT-6.1 Sol), and three leave significantly more: Gemini~3.8, 3.7, and 3.6 Flash. On DeepSWE (\figref{fig:deltagold}b), mean \(\NP_{\mathrm{code}}\) is 0.797, against 0.785 for the 103 developer patches; only two models are significantly below the developer baseline and none significantly above (mean gap \(-0.005\)). The sign of the SWE-bench Verified gap depends on the oracle: under \(\Oracle^{+}\) it flips from \(+2.3\) to \(-4.5\) percentage points (Appendix~\ref{app:oracle}).

\finding{1}{Agent patches are larger than developers' (2.33 times the lines). Outside tests, agents change 1.09 times the lines, yet most of their code is needed. On SWE-bench Verified, 11 of 18 models leave less unneeded code than developers (mean \(\NP_{\mathrm{code}}\) 0.887 vs.\ 0.864), four show no significant difference, and the three Gemini Flash models leave significantly more. On DeepSWE, agents are on par with developers (0.797 vs.\ 0.785), but no model is significantly above the developer baseline. On average, 11\% of an agent patch's code units are unneeded.}

\subsection{RQ2: What Are the Unneeded Edits?}
\label{sec:rq2}\label{sec:behavior}

\begin{figure}[t]
\vspace{-4pt}
\centering\includegraphics[width=0.5\linewidth,trim={8 10 10 10},clip]{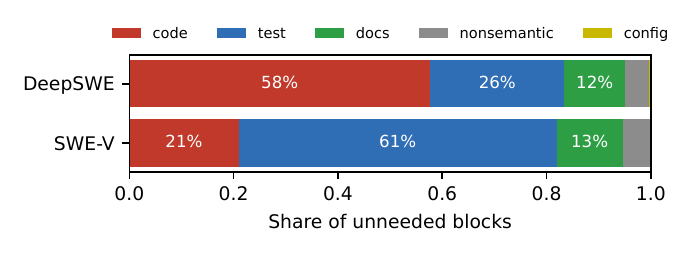}
\caption{Composition of extra edits: ungraded test edits and unneeded units (18 models; 6{,}728 SWE-bench Verified and 932 DeepSWE patches, all under \(\Oracle_{\mathrm{test}}\)). On DeepSWE, code is 58\% of extra edits.}
\label{fig:classes}
\end{figure}

\figref{fig:classes} shows the classes of extra edits. On SWE-bench Verified, 61\% are test-file edits, 21\% are code, 13\% are documentation, and 5\% are comment or formatting edits. Test edits are pervasive (81.5\% of agent patches edit a test file) and ungraded, and a reviewer may welcome many of them. Counting them as unneeded would dominate the all-unit \(\NP\), which then mostly tracks how many tests a model writes. On DeepSWE the composition flips: tasks are feature implementations with about six times more units per patch. Code is 58\% of extra edits, and tests are 26\% (Appendix~\ref{app:ext}).

On SWE-bench Verified, among the 3{,}987 unneeded code units, the behavioral probes (run on 3{,}559 unneeded agent units) show that 89\% change behavior, for example by raising or swallowing an exception on nearby inputs. The other 11\% are inert guards or duplicate checks (Appendix~\ref{app:oracle}).

We use mutation testing~\citep{jia2011mutation,just2014mutants} to ask whether the tests agents add are useful. Over the 271 SWE-bench Verified instances that pass our validity checks, we inject 1{,}962 mutants (small artificial bugs) into the patched code with \texttt{mutmut}~\citep{hovmoller2026mutmut}. The agent's added tests catch 68.1\% of them: more than the existing regression tests in the same files (45.6\%), but fewer than the task tests (77.1\%). They catch 73.2\% of mutants in needed code and 31.2\% in unneeded code. The agent's tests also catch bugs the task tests miss. Of the mutants that slip past the task tests, 22.7\% (102) are caught by tests the agent added or strengthened; this is 5.2\% of all mutants (Appendix~\ref{app:testmutation}). Most agent tests are written after the fix, and few show a failing-then-passing run (Appendix~\ref{app:ext}).

\finding{2}{On SWE-bench Verified, beyond the code, patches mostly add tests (61\%) and documentation (13\%); on DeepSWE, most extra edits are code (58\%). The added tests catch more injected bugs than the regression tests (68.1\% vs.\ 45.6\%), though fewer than the task tests (77.1\%), and they catch 22.7\% of the bugs the task tests miss, so agent-written tests can help strengthen benchmark test suites. Finally, 89\% of unneeded code units change behavior.}

\subsection{RQ3: Where Do Unneeded Edits Come From?}
\label{sec:rq3}

If residue came mostly from debugging, it would enter the patch at debug steps and never be reverted. The pooled debug share over all residue is \(\rho=0.257\), but 4{,}676 of the 4{,}898 debug-step residue units are test-file hunks that the agent wrote before the fix as regression tests. On code, only 154 of the 3{,}987 code-residue units (in 55 patches) come from debug steps, so \(\rho_{\mathrm{code}}=0.039\). Scratch scripts and debug prints almost never survive into resolved patches.

Fix steps write 83\% of unneeded code, and revision steps write 11\%. Code written while revising a fix is unneeded only slightly more often than code from the first fix (0.29 vs.\ 0.25), and revisions are rare (Appendix~\ref{app:ext}). On DeepSWE, over units with exact provenance, \(\rho_{\mathrm{code}}\) is 0.037.

\finding{3}{Debug steps write almost none of the unneeded code (154 of 3{,}987 units, \(\rho_{\mathrm{code}}=0.039\)); fix steps write 83\% of it, and code from revisions is unneeded only slightly more often than first-fix code (0.29 vs.\ 0.25).}

\subsection{RQ4: Which Models Over-Edit More?}
\label{sec:rq4}\label{sec:h2}

Across the 18 models on SWE-bench Verified, the Spearman correlation between resolve rate and \(\NP_{\mathrm{code}}\) is \(-0.10\) (raw \(p=0.69\); 95\% CI [\(-0.28\), \(+0.05\)], bootstrap over each model's instances). At nearly the same resolve rate, Sonnet~5 (80.6\%) has \(\NP_{\mathrm{code}}=0.922\) and Gemini~3.7 Flash (81.4\%) has 0.789, so Gemini~3.7 Flash leaves about 3.6 times as many unneeded code units per task (1.23 vs.\ 0.34).

The model effect survives controls for unit granularity and task selection (Appendix~\ref{app:rq4}). Multi-unit and line-weighted variants keep the ranking, and task fixed effects leave the same three Gemini Flash models below developers.

On DeepSWE the relation reverses: \(\NP_{\mathrm{code}}\) is negatively correlated with resolve rate (Spearman \(-0.59\), \(p=0.01\); \(-0.52\), \(p=0.03\), task-adjusted), so models that solve more tasks leave more unneeded code. The model ranking agrees moderately between benchmarks (\(+0.59\)), and Gemini~3.7 and 3.8 Flash are low on both (\tabref{tab:generalization}).

\finding{4}{The amount of unneeded code differs by model. It does not track resolve rate on SWE-bench Verified (Spearman \(-0.10\)), even after controlling for unit granularity and task selection, but on DeepSWE models with higher resolve rates leave more unneeded code (\(-0.59\)).}

\subsection{RQ5: Can Unneeded Code Be Removed Automatically?}
\label{sec:rq5}\label{sec:mechanism}

Task tests are unavailable at deployment, so we test three strategies that do not need them (\tabref{tab:cleanup}). All RQ5 experiments use SWE-bench Verified.

\paragraph{Stopping at the first green run.}
The numbers below cover all resolved SWE-bench Verified patches. For each trajectory we find the agent's first all-green run of its own tests, replay to that step, and grade the patch. The early patch still passes the task tests in 2{,}948 of 3{,}031 graded cases (97.3\%), but code units barely change (7{,}419 to 7{,}269) and \(\NP_{\mathrm{code}}\) on those patches moves only from 0.700 to 0.710.

\paragraph{Self-audit.}
We ask each model to justify every hunk of its final diff and drop the ones it cannot. Across all resolved SWE-bench Verified patches (17 of 18 models; Appendix~\ref{app:rq5}), dropped code hunks match a unit that \(\Oracle^{+}\) calls unneeded 65.7\% of the time, but the audit finds only 35.5\% of all such units (\(\kappa=0.39\)); the drops break 1.8\% of resolved patches. Models are selective: they drop 40.5\% of unneeded units that behave like the patch without them, against 14.7\% of those that behave like the full patch. Applying the drops raises \(\NP_{\mathrm{code}}\) from 0.924 to 0.944.

\paragraph{Pruning with \DRA{}.}
Most agents run their own tests, and 4{,}860 of 6{,}728 trajectories reach a green run. We run a simplified cumulative \DRA{} that reverts a unit whenever the agent's test command \(\Oracle_{\mathrm{agent}}\) still passes, and we also require the existing PASS\_TO\_PASS tests (\(\Oracle_{\mathrm{agent+P2P}}\)). This applies to 2{,}495 of the 6{,}724 patches, because the others have no standalone test command or fail the baseline check (Appendix~\ref{app:cleanup-eligibility}). With \(\Oracle_{\mathrm{agent+P2P}}\), pruning removes 67.3\% of the unneeded code units, nearly twice the self-audit's recall, but at 46.3\% precision against the self-audit's 65.7\%. It raises \(\NP_{\mathrm{code}}\) from 0.936 to 0.975, and 89.1\% of patches still pass the task tests. With \(\Oracle_{\mathrm{agent}}\) alone, pruning reaches 39.6\% precision and 66.5\% recall, and 86.6\% of its patches still pass the task tests. Most false drops are code units that implement behavior that neither the existing tests nor the agent's tests exercise.

\begin{table}[t]
\tablestyle{4pt}{1.15}
\small
\begin{tabular}{lcccc}
\toprule
& \textbf{First green} & \textbf{Self-audit} & \textbf{\DRA{} + \(\Oracle_{\mathrm{agent}}\)} & \textbf{\DRA{} + \(\Oracle_{\mathrm{agent+P2P}}\)} \\
\midrule
Patches / units graded                                 & 3{,}031 / 7{,}419   & 6{,}345 / 13{,}729 & 2{,}495 / 5{,}060 & 2{,}495 / 5{,}060 \\
Dropped units                                          & 150 (2.0\%, net)    & 1{,}238 (9.0\%) & 997 (19.7\%) & 863 (17.1\%) \\
Precision                                              & --          & \textbf{65.7\%} & 39.6\%   & 46.3\% \\
Recall                                                 & --           & 35.5\%   & 66.5\%   & \textbf{67.3\%} \\
\(\NP_{\mathrm{code}}\) before \(\to\) after           & 0.700 \(\to\) 0.710 (3\%) & 0.924 \(\to\) 0.944 (26\%) & 0.936 \(\to\) 0.974 (59\%) & 0.936 \(\to\) \textbf{0.975} (\textbf{61\%}) \\
Patches still resolving                                & 97.3\% & \textbf{98.2\%} & 86.6\%   & 89.1\% \\
\bottomrule
\end{tabular}
\caption{Post-generation cleanup on SWE-bench Verified under \(\Oracle^{+}\). Units are code units; dropped-unit percentages are of units graded, and \(\NP_{\mathrm{code}}\) is shown before and after reduction on each column's patches, with the share of the unneeded-code gap (\(1-\NP_{\mathrm{code}}\)) closed in parentheses. \(\Oracle_{\mathrm{agent}}\) is the agent's own test command; \(\Oracle_{\mathrm{agent+P2P}}\) also requires the existing PASS\_TO\_PASS regression tests. The \DRA{} columns cover the 2{,}495 eligible patches where \(\Oracle_{\mathrm{agent}}(S_T)=1\). First green covers the patches edited after it; its pass rate uses the task tests, and its units are the net change in code units. The self-audit covers 17 models; 1{,}540 patches had at least one drop, and the passing rate is on the regraded patches (none ungraded).}
\label{tab:cleanup}
\end{table}

\finding{5}{On SWE-bench Verified, self-audit and test-based pruning trade precision for recall. Self-audits are precise (65.7\%) but find only about a third of unneeded code (35.5\%), and their drops break 1.8\% of resolved patches. Pruning with the agent's own tests and the repository's regression tests finds about two thirds (67.3\%) at lower precision (46.3\%) on the 2{,}495 eligible patches, and 89.1\% of them still resolve. Early stopping barely changes the patch (\(\NP_{\mathrm{code}}\) 0.700 to 0.710), because most unneeded code is already in the patch when the agent's tests first pass.}

\section{Discussion and Limitations}
\label{sec:discussion}\label{sec:validity}\label{sec:limitations}\label{sec:future}

\paragraph{Does the metric measure what we claim?}
We check the metric against the developer patch. A code unit is \emph{outside-patch} if it lies outside every function the developer patch touched. Of 14{,}591 agent code units on 6{,}596 instances, 2{,}607 are outside-patch. The \DRA{} unneeded flag predicts outside-patch units with an AUROC of 0.718, against 0.478 for unit size (difference \(+0.24\), 95\% CI [0.21, 0.27]); it flags 3{,}753 units at precision 0.43 and recall 0.62. At the instance level, the unneeded-line fraction correlates with the outside-patch fraction at Spearman 0.40, against 0.21 for edit distance and 0.25 for lines relative to the developer patch. The metric flags code a developer would likely not have written better than patch size does, though fewer than half of flagged units are outside-patch. Whether a reviewer would remove the flagged units needs human labels (Appendix~\ref{app:ext}).

\paragraph{Implications.}
The amount of unneeded code a model writes does not track its resolve rate on SWE-bench Verified and is anti-correlated with it on DeepSWE, and most unneeded code changes behavior, so reviewers cannot skip it (\secref{sec:rq2}, \secref{sec:rq4}). Because code necessity differs by model, coding benchmarks should report \(\NP_{\mathrm{code}}\) with a developer-patch baseline alongside resolve rate. Size comparisons with developer patches should separate test and code edits. Debug steps write little code residue, so in-loop cleanup would have little to remove, although we did not test it, and post-generation cleanup is the more promising target. Its two forms trade off: self-audit rarely drops needed code but misses most unneeded code, while pruning with the agent's and the repository's regression tests finds most of it at the cost of more false drops (\secref{sec:rq5}). Because the unneeded code is written while constructing and revising the fix, we recommend debloating training trajectories: drop units that the tests do not need before supervised fine-tuning, in line with edit-aware training~\citep{yang2026paft,ke2026prepair} and as PDB proposes~\citep{chai2026pdb}.

\paragraph{Limitations.}
Necessity depends on the oracle, and on DeepSWE and SWE-bench Pro it is an upper bound on unneeded code (on DeepSWE the developer baseline is 0.785). Behavioral checks are LLM-written, and oracle-unneeded code can still be wanted. Units are reverted one at a time, so jointly needed pairs and mixed blocks can mislead. We study over-editing in successful trajectories, i.e., resolved patches, and leave over-editing in unsuccessful trajectories to future work. Role labels were validated against a second LLM judge only, with no human audit (Appendix~\ref{app:roles}). The debug-share result rests on 154 of 3{,}987 code-residue units, and the mutation results cover 271 instances (1{,}962 mutants). Provenance needs replayable trajectories: 3.3\% of SWE-bench Verified replays are not exact (228 replay mismatches, 6 self-check failures, 5 binary patches), and 33.4\% of DeepSWE patches (311 of 932) contain at least one unit without provenance (14.9\% of units, 12.7\% of code units). The main results use one harness, Copilot CLI (Appendix~\ref{app:ext} covers others). SWE-bench Verified is Python-only and may be contaminated.

\paragraph{Future work.}
Human labels (Appendix~\ref{app:annotation}) would test whether reviewers agree with the oracle. Property-based testing~\citep{maciver2019hypothesis} and LLM-generated differential tests~\citep{hu2026difftestgen} could extend probe-based grading to benchmarks without developer patches. Repeating the experiment on open-source harnesses such as OpenHands~\citep{wang2025openhands} and mini-SWE-agent~\citep{minisweagent}, using public trajectories~\citep{swebenchexperiments}, would show how harness design affects over-editing.

\section{Conclusion}
\label{sec:conclusion}

We took 7{,}660 resolved coding-agent trajectories and traced each block of the final patch to the step that wrote it, then graded necessity with the task tests and validated unneeded code with behavioral probes. Agent patches are larger than developers', but on SWE-bench Verified the extra lines are mostly tests and documentation. On code, on average about 11\% of an agent patch's code units are unneeded; on SWE-bench Verified, 11 of 18 models leave less unneeded code than developers and the three Gemini Flash models leave more, while on DeepSWE agents are on par with developers. The unneeded code that remains is written while building and revising the fix, usually changes behavior, and differs by model, with no relation to resolve rate on SWE-bench Verified and a negative one on DeepSWE. After generation, self-audit removes it precisely but finds only about a third, while pruning with the agent's own tests and the repository's regression tests finds about two thirds at lower precision on the patches where it applies. We recommend reporting code necessity next to resolve rate and addressing over-editing through model training and harness-level pruning.
\label{endmain}

\clearpage
\bibliographystyle{plainnat}
\bibliography{references}

\newpage
\appendix
\makeatletter
\let\agentlensorigsection\section
\let\agentlensorigsubsection\subsection
\RenewDocumentCommand{\section}{s o m}{%
  \IfBooleanTF{#1}
    {\agentlensorigsection*{#3}}%
    {%
      \IfNoValueTF{#2}{\agentlensorigsection{#3}}{\agentlensorigsection[#2]{#3}}%
      \addcontentsline{atoc}{section}{\protect\numberline{\thesection}\IfNoValueTF{#2}{#3}{#2}}%
    }%
}
\RenewDocumentCommand{\subsection}{s o m}{%
  \IfBooleanTF{#1}
    {\agentlensorigsubsection*{#3}}%
    {%
      \IfNoValueTF{#2}{\agentlensorigsubsection{#3}}{\agentlensorigsubsection[#2]{#3}}%
      \addcontentsline{atoc}{subsection}{\protect\numberline{\thesubsection}\IfNoValueTF{#2}{#3}{#2}}%
    }%
}
\newcommand{\agentlensappendixtocentry}[4]{%
  \par
  \addvspace{#1}%
  \noindent
  \begingroup
    \parindent=0pt
    \rightskip=0pt plus 1fil
    \parfillskip=0pt
    \leavevmode
    \hspace*{#2}{#3}%
    \nobreak\leaders\hbox to 0.65em{\hss.\hss}\hfill
    \nobreak\hb@xt@ 2.5em{\hfil\color{black}#4}\par
  \endgroup
}
\newcommand{\agentlensappendixsectiontoc}[2]{%
  \begingroup
  \def\numberline##1{\makebox[1.8em][l]{##1}}%
  \agentlensappendixtocentry{0.12em}{0pt}{\bfseries #1}{#2}%
  \endgroup
}
\newcommand{\agentlensappendixsubsectiontoc}[2]{%
  \begingroup
  \def\numberline##1{\makebox[3.2em][l]{##1}}%
  \agentlensappendixtocentry{-0.10em}{2.2em}{#1}{#2}%
  \endgroup
}
\makeatother

\setlength{\textfloatsep}{8pt plus 2pt minus 2pt}
\setlength{\floatsep}{6pt plus 2pt minus 2pt}
\setlength{\intextsep}{6pt plus 2pt minus 2pt}

\section*{Appendix}

{\bfseries Table of Contents\par}
\vspace{-1.3em}
\noindent\rule{\linewidth}{0.5pt}

{
\setcounter{tocdepth}{2}
\makeatletter
\begingroup
\let\l@section\agentlensappendixsectiontoc
\let\l@subsection\agentlensappendixsubsectiontoc
\@starttoc{atoc}
\endgroup
\makeatother
}

\newpage


\section{Pre-Registered Hypotheses and Deviations}
\label{app:hyp-group}

\subsection{Pre-Registered Hypotheses}
\label{app:hypotheses}

Before the main experiments we registered five hypotheses with support and kill criteria. A two-model pilot (Opus~5 and Haiku~4.5, 50 instances each) met the registered bar for continuing (pooled \(\rho\ge 0.15\)). \tabref{tab:hypotheses} lists the hypotheses and \tabref{tab:hyp-outcome} summarizes what each study found. The main text reports the studies as research questions. The hypotheses keep their registered wording, in which \emph{residue} meant unneeded units written at debug steps; the main text calls all unneeded units residue and reports the debug share \(\rho\).

\begin{table}[H]
\tablestyle{3pt}{1.15}
\begin{tabular}{lly{105}y{85}y{85}}
\toprule
\textbf{H} & \textbf{RQ} & \textbf{Hypothesis} & \textbf{Support if} & \textbf{Kill or redirect if} \\
\midrule
H1 & RQ3 & Residue dominance: much of the unneeded code in resolved patches is debug residue. & Pooled \(\rho \ge 0.40\). & \(\rho < 0.15\): unneeded code is in the fix; redirect to training-time causes. \\
H2 & RQ4 & Second axis: \(\NP\) separates models that resolve rate does not. & Spearman(resolve, \(\NP\)) \(<0.5\) and at least one pair within 3\,pp with non-overlapping CIs. & Spearman \(\ge 0.8\). \\
H3 & Validity & Metric validity: \(\NP\) tracks human ``should remove'' labels better than edit-distance. & AUROC gain \(\ge 0.1\). & No advantage. \\
H4 & RQ5 & Harness fix: revert-on-reject plus a diff audit cut \(|E|\) by \(\ge 20\%\). & Resolve rate within 1\,pp. & Resolve loss \(>1\)\,pp. \\
H5 & RQ3 & Preservation prompts act on \(\Efix\), not \(\Eres\). & \(|\Efix|\) drops \(\ge 20\%\) but \(|\Eres|\) drops \(<10\%\). & Both drop alike. \\
\bottomrule
\end{tabular}
\caption{Pre-registered hypotheses.}
\label{tab:hypotheses}
\end{table}

\begin{table}[H]
\tablestyle{3pt}{1.15}
\begin{tabular}{ly{70}y{270}}
\toprule
\textbf{H} & \textbf{Status} & \textbf{What the data show} \\
\midrule
H1 & Depends on statistic & Pooled \(\rho=0.257\) over all units falls between the kill (0.15) and support (0.40) thresholds. Almost all of it is regression-test hunks; over code units \(\rho_{\mathrm{code}}=0.039\), below the kill threshold. The conclusion therefore rests on switching to \(\rho_{\mathrm{code}}\). \\
H2 & Met, with a changed statistic & The registered criteria used the all-unit \(\NP\) (Spearman \(+0.069\)); the main text uses \(\NP_{\mathrm{code}}\) (\(-0.10\), 19 separated pairs). Both pass. \\
H3 & Stage 1 only & Only the developer-patch proxy (stage 1) was run: AUROC gain \(+0.24\) (0.718 vs.\ 0.478), above the 0.1 bar on the proxy. The human stage has LLM pre-labels (\(\kappa=0.63\)) and no human labels, so H3 is not confirmed. \\
H4 & Deferred & Not run; it targets debug-step edits, which leave little code residue. RQ5 tests offline proxies. \\
H5 & Deferred & Not run, for the same reason. \\
\bottomrule
\end{tabular}
\caption{Outcome of each pre-registered hypothesis.}
\label{tab:hyp-outcome}
\end{table}

\subsection{Deviations from the Registered Plan}
\label{app:deviations}

\tabref{tab:deviations} lists where the study departs from the registered plan and why.

\begin{table}[H]
\tablestyle{4pt}{1.2}
\begin{tabular}{y{110}y{140}y{150}}
\toprule
\textbf{Registered} & \textbf{Done} & \textbf{Reason and effect} \\
\midrule
Statement-level AST units with dependency closure (GumTree). & Contiguous line blocks; new file = one unit. & Language-independent and exactly revertible. Coarser units make single-unit patches more common; the multi-unit and line-weighted checks address this (\secref{sec:rq4}). \\
Strengthened oracle with property-based differential testing (\textsc{Live}/\textsc{Inert}). & Strengthened oracle \(\Oracle^{+}\) combining \(\Oracle_{\mathrm{test}}\) with 8 LLM-written differential behavioral probes per test-unneeded code unit (\secref{sec:oracle}; Appendix~\ref{app:oracle}). & Property-based testing is future work (\secref{sec:discussion}). \(\Oracle^{+}\) counts developer-equivalent untested code as needed and splits remaining unneeded units into live vs.\ inert. \\
Debug share \(\rho\) over all units as the H1 statistic. & Reported, plus \(\rho_{\mathrm{code}}\) over code units. & The pilot showed most extra edits are test-file edits, which the test oracle cannot grade. \\
Agent \(\NP\) compared with the developer baseline. & Matched \(\Ddev\) on \(\NP_{\mathrm{code}}\) as the main comparison. & Developer patches have no test-file units in the oracle's sense; comparing code with code is like for like. \\
All resolved SWE-bench Verified instances per model. & All resolved instances for RQ1--RQ5, the cleanup analyses, the strengthened oracle, and the mutation analysis (eligible frame). & Grading cost for the probe-based analyses; task fixed effects and the 99-task paired comparison control for task mix. \\
Single role/probe judge. & Role and probe judges are GPT-6.1 Sol. & Roles were validated against a second judge (Appendix~\ref{app:roles}). \\
Role judge on every step; hand audit of 150 steps. & Rules plus judge on unit-introducing steps; second-judge check on 10{,}000 sampled steps. & Only introducing steps enter \(\rho\). No hand audit was done. \\
Generalization on mini-SWE-agent and OpenHands trajectories. & Final-patch \(\NP\) under four other harnesses; no public trajectories. & Run access; roles are not generalized beyond Copilot CLI. \\
Exploratory E1 (debug share vs.\ trajectory length) and E2 (review-aware routing). & Not run. & Debug steps write too little code residue for E1 to be informative; E2 is left to follow-up work. \\
H4 and H5 intervention runs. & Deferred; offline self-audit as a proxy for the diff-audit intervention. & They target debug-step edits, which H1 shows leave little code residue. \\
\bottomrule
\end{tabular}
\caption{Deviations from the registered plan.}
\label{tab:deviations}
\end{table}

\section{Method Details}
\label{app:method-group}

\subsection{Running Example}
\label{app:example}

\noindent\begin{minipage}{\linewidth}
Consider an issue reporting that \texttt{parse\_date("")} raises \texttt{ValueError} instead of returning \texttt{None}; the task's test asserts \texttt{parse\_date("") is None}. The developer fix adds two lines. Suppose an agent submits the following patch, whose four units entered at different steps:
\begin{lstlisting}[basicstyle=\ttfamily\footnotesize, columns=fullflexible, keepspaces=true, frame=lines, xleftmargin=3em, xrightmargin=3em, aboveskip=6pt, belowskip=4pt]
 def parse_date(s):
+    if not isinstance(s, str):          # u1   t3 Probe
+        return None
+    log.debug("parse_date(%r)", s)      # u4   t4 Probe
+    if s == "":                         # u2   t5 Fix
+        return None
-    d = datetime.strptime(s, FMT)
-    return d.date()
+    dt = datetime.strptime(s, FMT)      # u3   t6 Refine
+    return dt.date()
\end{lstlisting}
\end{minipage}

\noindent After exploring (\(t_1\)) and reproducing the error (\(t_2\)), the agent adds a type guard \(u_1\) (\(t_3\), \role{Probe}) and a debug log \(u_4\) (\(t_4\), \role{Probe}). The fix \(u_2\) follows (\(t_5\), \role{Fix}), and a final pass renames a local variable (\(u_3\), \(t_6\), \role{Refine}). The reproduction script is deleted at \(t_7\) and leaves nothing in \(\Delta\). Reverting \(u_2\) fails the task's test, so \(u_2\) is needed. Reverting \(u_1\), \(u_3\), or \(u_4\) keeps the tests passing, so \(\NP=1/4\) and the unneeded units are \(\{u_1,u_3,u_4\}\). These three are the patch's residue; two were introduced at \role{Probe} steps, so \DRA{} attributes them to debugging. RQ3 shows this pattern is rare: in resolved patches, unneeded code looks like \(u_3\), not like \(u_1\) or \(u_4\).

\subsection{Three Notions of Minimality}
\label{app:minimality}

\tabref{tab:minimality} contrasts the minimality notions discussed in \secref{sec:prelim}.

\begin{table}[H]
\tablestyle{4pt}{1.2}
\begin{tabular}{y{80}y{130}y{110}y{110}}
\toprule
\textbf{Notion} & \textbf{Definition} & \textbf{Cost to establish} & \textbf{Main weakness} \\
\midrule
Global minimum & Fewest units among all patches accepted by \(\Oracle\). & Search over subsets of candidate edits; exponential in general~\citep{zeller2002dd}. & Infeasible at repository scale; still oracle-relative. \\
1-minimal & Removing any single unit makes \(\Oracle\) reject. & Verification: \(|\Units(\Delta)|\) oracle calls; reduction via delta debugging (ddmin) and related reducers~\citep{zeller2002dd,misherghi2006hdd,sun2018perses}. & Inherits oracle incompleteness; sensitive to unit granularity and interactions. \\
Semantic minimum & Smallest semantic change satisfying a specification, e.g., via MaxSMT~\citep{mechtaev2015directfix}. & Requires a specification (often tests) and constraint solving. & Specifications are rarely available; tests used as specifications overfit~\citep{smith2015cure}. \\
\bottomrule
\end{tabular}
\caption{Notions of patch minimality. \DRA uses 1-minimality with respect to the task tests as a tractable, explicitly oracle-relative measure.}
\label{tab:minimality}
\end{table}

\(\NP\) establishes that each unit is required by a stated oracle, so reporting the oracle, and a developer baseline graded by the same oracle, alongside \(\NP\) is part of the metric.

\subsection{Replay, Units, and Grading Details}
\label{app:replay}

\paragraph{Replay.}
For each instance we fetch the base version of every file the final patch touches and apply the trajectory's file-editing tool calls (\texttt{edit} string replacements, \texttt{create}, and patch-application calls) in order, keeping for each line of each file the step that last wrote it and, for each deleted base line, the step that deleted it. When a file was written by a shell command (for example \texttt{sed} or a heredoc), we attribute it to the last shell command that writes and names that file and reconstruct the file state after it by undoing the later structured edits; if that is not possible, the file's units keep their necessity grade but have no provenance. A trajectory is kept only if the replayed final contents equal the submitted patch for every file.

\paragraph{Units.}
For each file we diff base and final contents with Python's \texttt{difflib} (no autojunk) and merge change blocks separated by at most one unchanged line. A created or deleted file is one unit. Units are classified by path and content, in this order. \emph{Test}: the file is in the official test patch, or its path is in a \texttt{test}/\texttt{tests}/\texttt{testing}/\texttt{\_\_tests\_\_}/\texttt{testdata}/\texttt{e2e} directory or matches a test-file pattern (\texttt{test\_*.py}, \texttt{*\_test(s).py}, \texttt{conftest.py}, \texttt{*\_test.go}, \texttt{*.test/spec.[jt]s(x)}, snapshots, \texttt{tests.rs}). \emph{Docs}: the path is under \texttt{doc}/\texttt{docs}, ends in \texttt{.rst}, \texttt{.txt}, or \texttt{.md}, or is a changelog, authors, or news file. \emph{Config}: JSON, YAML, TOML, lock, module, and similar files. \emph{Nonsemantic}: a Python file whose AST, with docstrings removed, is unchanged by the unit, or a C-like source file whose token sequence without comments and whitespace is unchanged. \emph{Code}: everything else. Developer patches, SWE-bench Pro patches, and other-harness patches are unitized the same way, without replay.

\paragraph{Grading.}
For unit \(k\), the variant patch is the final patch with unit \(k\) restored to its base lines (or the file removed or restored). Each variant is graded with the task's FAIL\_TO\_PASS and PASS\_TO\_PASS tests in the official evaluation image, together with a control run of the full patch; instances whose control run fails are excluded. Units whose reversion would leave the base state are recorded as necessary without a run (742 units on SWE-bench Verified); reversions that do not parse are counted as necessary (39 units); and variants whose test run hung repeatedly are recorded as failures, i.e., necessary (22 units). An evaluation status of \texttt{error} counts as unresolved.

\subsection{Step-Role Rules and Judge Prompt}
\label{app:roles}

\tabref{tab:rolerubric} defines the step roles.

\begin{table}[H]
\tablestyle{4pt}{1.2}
\begin{tabular}{y{70}y{330}}
\toprule
\textbf{Role} & \textbf{Description} \\
\midrule
\role{Explore} & Reads, searches, or lists code without modifying tracked files. \\
\role{Reproduce} & Creates or edits a script or test whose purpose is to reproduce or check the issue. Rule: a newly created file that is executed (\texttt{python}, \texttt{pytest}, \texttt{node}, \texttt{bash}) within the next three steps. \\
\role{Probe} & Inserts instrumentation or tries a change to test a hypothesis. Rule: an edit whose added lines are all print, logging, \texttt{pdb}/\texttt{breakpoint}, or \texttt{console.log} statements. \\
\role{Fix} & Applies the change the agent presents as its solution. \\
\role{Refine} & Modifies an existing fix after test or self-review feedback. \\
\role{Cleanup} & Reverts or deletes earlier changes. \\
\bottomrule
\end{tabular}
\caption{Step roles. Only unit-introducing steps are labeled. The two rules run first; the judge labels the remaining steps.}
\label{tab:rolerubric}
\end{table}

The judge (GPT-6.1 Sol, no tools) receives the following prompt; debug roles are \role{Explore}, \role{Reproduce}, and \role{Probe}.
\begin{lstlisting}
You are labeling one step of a software-engineering agent trajectory.
Issue summary: {issue}
Previous steps (condensed): {history}
Current action: {action}
Observation (truncated): {observation}
Working-tree change caused by this step: {diff_delta}
Agent narration in the following steps: {after}
Agent's final message describing its solution: {final}

Choose exactly one role from:
EXPLORE, REPRODUCE, PROBE, FIX, REFINE, CLEANUP.
Use FIX only if the agent treats this change as part of its intended solution.
Use PROBE if the change is made to test a hypothesis or gather information.
Use REFINE if the step modifies an existing fix after test or self-review feedback.
Use REPRODUCE if the step creates or edits a script/test whose purpose is to
reproduce or check the issue.
Answer with the role and one sentence of justification, formatted as:
ROLE: <role>
WHY: <sentence>
\end{lstlisting}

\paragraph{Judge agreement.}
Role labels were validated against a second LLM judge (Claude Opus 5.5) only; no human audit exists. On a uniform random sample of 10{,}000 judge-labeled steps, the second judge agrees with the primary on 84.4\% of role labels (\(\kappa=0.69\); per-model \(\kappa\) 0.59--0.86) and on 89.1\% of debug vs.\ non-debug role labels (\(\kappa=0.55\)). Agreement is 82.1\% (\(\kappa=0.65\); debug 87.0\%, \(\kappa=0.56\)) on SWE-bench Verified and 88.4\% (\(\kappa=0.76\); debug 93.1\%, \(\kappa=0.50\)) on DeepSWE. The most common disagreement is \role{Reproduce} labeled \role{Fix} by the second judge (786 steps), a direction that would shrink \(\rho\), so the primary labels are conservative for the debug-share finding.

\subsection{Human Validation}
\label{app:annotation}

This appendix gives the protocol for the human labeling stage of the metric-validity check, which we leave to future work; the packet and the LLM pre-labels are prepared. The packet has 80 code units: 20 from developer patches, 30 from the two lowest-\(\NP_{\mathrm{code}}\) models, and 30 from high-\(\NP_{\mathrm{code}}\) models, each half \DRA-needed and half unneeded; untested single-unit patches are excluded. Annotators see the issue text and the full patch with the target unit highlighted, but not \DRA's label, the step role, or the model identity. They label each unit \emph{required}, \emph{over-editing}, or \emph{unsure}, with a one-sentence reason. Labeling is LLM-assisted: two LLM judges (Claude Opus~5.5 and GPT-6 Astra) first label each unit from the same information, and each author approves or overrides their label, or labels directly where the judges disagree, recording a reason for every override. The two judges agree on 72 of 80 units (\(\kappa=0.63\)). Both judge models are in the evaluated pool, and annotators may anchor on the suggested label, so raw judge labels, human-final labels, and the override rate are to be reported separately. If agreement between authors is below \(\kappa=0.6\), or the CI of the AUROC difference straddles 0.1, the sample is expanded to 200 units with a third adjudicating annotator.

\section{Oracles}
\label{app:oracle-group}

\subsection{Strengthened Test Oracle and Behavioral Probes}
\label{app:oracle}

Benchmark test suites do not exercise every path a developer fix touches~\citep{wang2026patchdiff}. The primary measure is \(\NP\) under \(\Oracle_{\mathrm{test}}\) (developer baseline \(\NP_{\mathrm{code}}=0.864\) over 1{,}317 code units), and behavioral probes validate the unneeded verdicts. The strengthened oracle \(\Oracle^{+}\) uses the GPT-6.1 Sol probe judge. We probed 3{,}559 \(\Oracle_{\mathrm{test}}\)-unneeded agent code units and 432 of the 434 \(\Oracle_{\mathrm{test}}\)-unneeded developer code units (26 agent units on 9 patches lie on tasks without a developer source patch and are unprobed). The pooled \(\NP\) comparisons under \(\Oracle^{+}\) below use the probe cohort of 18 models, 6{,}728 patches and 482 unique tasks (469 with developer source patches), where \(\Oracle_{\mathrm{test}}\) gives an agent mean \(\NP_{\mathrm{code}}=0.887\) against a developer baseline of 0.864 (\(\Ddev=+0.023\), \(p=0.016\)). To separate untested but developer-equivalent fixes from truly unneeded code, we use differential behavioral probes inspired by PatchDiff~\citep{wang2026patchdiff}.

\paragraph{Probe results (GPT-6.1 Sol).}
GPT-6.1 Sol wrote the probes. Of 3{,}559 probed unneeded agent code units, 3{,}151 (89\%, 95\% CI [87, 90]) change behavior when reverted, against 363 of 432 (84\% [80, 87]) for developer code units. On 388 units judged by both, Opus~5.5 finds 81\% changing behavior and GPT-6.1 Sol 89\% (\(\kappa=0.53\); for developer units 70\% vs.\ 80\%, \(\kappa=0.64\)), so the judge change makes the probe stricter about inertness. Class agreement between judges is 93\% (287 of 308).

\paragraph{Probe protocol and breakdown.}
For each \(\Oracle_{\mathrm{test}}\)-unneeded unit \(u\) of a final patch \(S_T\), the probe judge reads the issue and the diff between \(S_T\) and \(S_T \revert u\) (without seeing the hidden tests or the developer patch) and writes a probe script that calls the changed code and prints labeled outputs. Each probe runs on \(S_T\), \(S_T \revert u\), and the developer-patched repository \(S_{\mathrm{dev}}\). On the 112 \(\Oracle_{\mathrm{test}}\)-needed agent code units used as a positive control, the probes detect a behavioral change on all 112 (100\%; 119 of 119 with developer units, \tabref{tab:behavior}). Among the 3{,}559 \(\Oracle_{\mathrm{test}}\)-unneeded agent units, 3{,}151 (88.5\%) change behavior when reverted: on 1{,}263 (35.5\%), \(S_{\mathrm{dev}}\) matches \(S_T\) rather than \(S_T \revert u\) (untested developer-aligned fix, needed under \(\Oracle^{+}\)); on 1{,}586 (44.6\%), \(S_{\mathrm{dev}}\) matches \(S_T \revert u\) (unrequested behavioral change); on 133 (3.7\%) the evidence is mixed and on 169 (4.7\%) \(S_{\mathrm{dev}}\) matches neither version. The remaining 408 (11.5\%) are inert on all probes. Under \(\Oracle^{+}\), the 2{,}296 remaining unneeded agent code units therefore split into 1{,}888 live divergences (82\%) and 408 inert edits (18\%). Replacing \(\Oracle_{\mathrm{test}}\) with \(\Oracle^{+}\) raises developer \(\NP_{\mathrm{code}}\) from 0.864 to 0.970 and mean agent \(\NP_{\mathrm{code}}\) from 0.887 to 0.925, so the developer-baseline gap flips from \(+2.3\) to \(-4.5\) percentage points (\(p=9.6\times10^{-5}\); \(-4.8\pm0.7\) with task fixed effects, \(p=4\times10^{-11}\); \(-11.0\) on patches with at least two code units, \(p=1.1\times10^{-9}\)). Under the live-behavior oracle \(\Oracle_{\mathrm{live}}\) (any behavior change counts as needed) agents are \(+0.8\) points above developers (0.978 vs.\ 0.970, \(p=2.4\times10^{-4}\)). Per model, \(\Oracle_{\mathrm{live}}\) puts 14 of 18 models above developers and one below, while \(\Oracle^{+}\) puts 12 models below developers and none above. The 428 unneeded agent units that could not be probed (234 have no developer code, 194 timed out) do not change the sign: restricted to patches with all units probed, the \(\Oracle_{\mathrm{live}}\) and \(\Oracle^{+}\) gaps are \(+1.5\) and \(-3.8\) points. \tabref{tab:results_otest} reports the per-model results with 95\% bootstrap confidence intervals under the benchmark task tests (\(\Oracle_{\mathrm{test}}\)) alone.

\begin{table}[H]
\tablestyle{5pt}{1.15}
\begin{tabular}{lrl}
\toprule
\textbf{Units (\(\Oracle_{\mathrm{test}}\) pre-filter)} & \textbf{Checked} & \textbf{Changes behavior} \\
\midrule
Needed agent units (positive control) & 112 & 112 (100\%) \\
Test-unneeded agent units & 3{,}559 & 3{,}151 (88.5\%) [87, 90] \\
Test-unneeded developer units & 432 & 363 (84.0\%) [80, 87] \\
\midrule
\multicolumn{3}{l}{\emph{Test-unneeded agent units, by how the developer patch behaves on the differing probes}} \\
\quad Like the full agent patch (needed under \(\Oracle^{+}\)) & 3{,}559 & 1{,}263 (35.5\%) [34, 37] \\
\quad Like the patch without the unit (unrequested change; unneeded under \(\Oracle^{+}\)) & 3{,}559 & 1{,}586 (44.6\%) [43, 46] \\
\quad Mixed (unneeded under \(\Oracle^{+}\)) & 3{,}559 & 133 (3.7\%) \\
\quad Neither (divergent change; unneeded under \(\Oracle^{+}\)) & 3{,}559 & 169 (4.7\%) \\
\quad No probe difference (inert; unneeded under \(\Oracle^{+}\)) & 3{,}559 & 408 (11.5\%) \\
\bottomrule
\end{tabular}
\caption{Differential behavior probes on \(\Oracle_{\mathrm{test}}\)-unneeded code units (SWE-bench Verified, GPT-6.1 Sol probe judge). Brackets are 95\% Wilson intervals.}
\label{tab:behavior}
\end{table}

\begin{table}[H]
\tablestyle{4.5pt}{1.12}
\begin{tabular}{lrcccc}
\toprule
\textbf{Model} & \textbf{Tasks} & \(\NP_{\mathrm{test}}\) & \(\NP_{\mathrm{code}}\) [95\% CI] & \(\Ddev\) vs.\ developer [95\% CI] & \(\rho\) \\
\midrule
\baseline{Developer patch} & \baseline{470} & \baseline{0.864$^{*}$} & \baseline{0.864} & \baseline{--} & \baseline{n/a} \\
Claude Opus 5.5 & 424 & 0.451 & 0.900 [0.879, 0.919] & \textbf{+0.042} [+0.018, +0.065] & 0.39 \\
Kimi K3 & 410 & 0.479 & 0.897 [0.876, 0.917] & +0.024 [+0.000, +0.047] & 0.36 \\
Claude Opus 5 & 420 & 0.657 & 0.898 [0.877, 0.919] & \textbf{+0.031} [+0.009, +0.055] & 0.25 \\
Claude Opus 4.8 & 404 & 0.657 & 0.912 [0.892, 0.930] & \textbf{+0.045} [+0.022, +0.069] & 0.34 \\
GPT-6 Astra & 400 & 0.279 & 0.889 [0.867, 0.910] & \textbf{+0.025} [+0.002, +0.047] & 0.24 \\
GPT-5.6 Sol & 388 & 0.422 & 0.924 [0.904, 0.942] & \textbf{+0.042} [+0.021, +0.063] & 0.02 \\
Gemini 3.8 Flash & 392 & 0.339 & 0.765 [0.735, 0.793] & \textbf{$-$0.095} [$-$0.129, $-$0.064] & 0.37 \\
Gemini 3.7 Flash & 392 & 0.348 & 0.789 [0.759, 0.820] & \textbf{$-$0.074} [$-$0.107, $-$0.042] & 0.39 \\
Claude Sonnet 5 & 382 & 0.634 & 0.922 [0.903, 0.942] & \textbf{+0.052} [+0.028, +0.077] & 0.34 \\
Claude Haiku 5.5 & 351 & 0.552 & 0.913 [0.891, 0.933] & \textbf{+0.044} [+0.016, +0.071] & 0.42 \\
GPT-6.1 Sol & 377 & 0.273 & 0.887 [0.865, 0.908] & +0.024 [$-$0.003, +0.048] & 0.18 \\
GPT-6 Sol & 363 & 0.333 & 0.886 [0.862, 0.908] & +0.011 [$-$0.016, +0.038] & 0.12 \\
GPT-5.6 Terra & 360 & 0.433 & 0.928 [0.910, 0.946] & \textbf{+0.058} [+0.035, +0.083] & 0.04 \\
Claude Sonnet 5.5 & 367 & 0.848 & 0.929 [0.911, 0.947] & \textbf{+0.055} [+0.033, +0.078] & 0.07 \\
GPT-5.6 Luna & 330 & 0.463 & 0.946 [0.928, 0.962] & \textbf{+0.059} [+0.035, +0.084] & 0.05 \\
Gemini 3.6 Flash & 349 & 0.374 & 0.791 [0.760, 0.820] & \textbf{$-$0.076} [$-$0.111, $-$0.041] & 0.38 \\
GPT-6 Luna & 312 & 0.431 & 0.921 [0.900, 0.941] & \textbf{+0.048} [+0.022, +0.072] & 0.08 \\
Claude Haiku 4.5 & 307 & 0.522 & 0.873 [0.847, 0.899] & $-$0.003 [$-$0.035, +0.030] & 0.38 \\
\midrule
All models & 6{,}728 patches & 0.472 & 0.887 & -- & 0.257 \\
\bottomrule
\end{tabular}
\caption{Per-model necessity on SWE-bench Verified (all resolved instances) under the benchmark task tests (\(\Oracle_{\mathrm{test}}\)), the primary oracle. \(^{*}\)Developer \(\NP_{\mathrm{test}}\) over all units is not reported separately; the developer value shown is \(\NP_{\mathrm{code}}\). Tasks is the number of tasks analyzed per model. Bold \(\Ddev\) indicates a 95\% instance-bootstrap CI excluding 0. \(\NP_{\mathrm{test}}\) CIs are not reported.}
\label{tab:results_otest}
\end{table}

\section{Extended Analyses}
\label{app:ext}

\subsection{RQ1 Details}
\label{app:rq1}

The median model changes 2.33 times as many lines as the developer's source patch (range 0.94--5.10 across models). Counting only non-test files, the median ratio is 1.09 (0.79--2.03). Counting only code units, it is 0.86 (0.72 for Sonnet~5.5 to 1.36 for Gemini~3.7 Flash), and the per-task median ratio is 1.00. The mean all-unit \(\NP\) under \(\Oracle_{\mathrm{test}}\) is 0.472. Under \(\Oracle_{\mathrm{test}}\), 33\% of developer code units (434 of 1{,}317) are unneeded; strengthening with behavioral checks reclassified 84\% of test-unneeded developer code units (363 of 432 probed) and 35\% of test-unneeded agent code units (1{,}263 of 3{,}559) as needed, leaving 3\% of developer code units unneeded (Appendix~\ref{app:oracle}). The three models significantly below developers have \(\Ddev\) of \(-0.095\) (Gemini~3.8 Flash), \(-0.074\) (Gemini~3.7 Flash), and \(-0.076\) (Gemini~3.6 Flash).

\subsection{RQ2 Details}
\label{app:rq2}

Besides tests (61\%), code (21\%), and documentation (13\%), 5\% of extra edits are nonsemantic edits (comments, docstrings, formatting). Per-model all-unit \(\NP\), which counts test edits as unneeded, ranges from 0.273 (GPT-6.1 Sol) to 0.848 (Sonnet~5.5), largely tracking how many tests a model writes; 60\% of all agent-changed lines are in test files. On DeepSWE, tasks are feature implementations with about six times more units per patch; code is 58\% of extra edits, tests 26\%, and documentation 12\%.

\paragraph{Added tests.}
Among the 5{,}439 SWE-bench Verified patches that edit tests and have step provenance (11{,}526 test units; 5{,}482 patches and 11{,}640 units overall), 67.2\% of test units reference symbols from needed code units, 52.0\% of these patches write all surviving tests after the last needed code edit, and 8.7\% see a test fail before the fix and pass after (Appendix~\ref{app:testmutation}).

\paragraph{Mutation testing.}
We evaluated 1{,}962 mutants in 271 instances (387 instances are eligible and 271 yielded valid mutants) with \texttt{mutmut} (v3.8.1)~\citep{hovmoller2026mutmut} under four suites: the existing regression tests in the touched test files, the agent's added tests, the task tests \(\Oracle_{\mathrm{test}}\), and \(\Oracle^{+}\). The agent's added tests detect 68.1\% of mutants, the regression tests 45.6\%, and \(\Oracle_{\mathrm{test}}\) 77.1\%. They detect 73.2\% of mutants in needed code and 31.2\% in unneeded code (\(\Oracle_{\mathrm{test}}\): 83.4\% and 31.2\%). Of the mutants \(\Oracle_{\mathrm{test}}\) misses, 22.7\% (102, or 5.2\% of all mutants) are caught by tests the agent added or strengthened. \(\Oracle^{+}\) detects 84.7\% (79.4\% under the strict variant), and the agent's added or strengthened tests catch a further 4.1\% of all mutants (81) that \(\Oracle^{+}\) misses. Appendix~\ref{app:testmutation} gives details.

\subsection{RQ3 Details}
\label{app:rq3}

Per-model \(\rho\) ranges from 0.018 (GPT-5.6 Sol) to 0.422 (Haiku~5.5), mostly reflecting whether a model writes tests first. Counting \role{Refine} as a debug role raises \(\rho_{\mathrm{code}}\) to 0.148. The 154 residue code units written at debug steps come from 55 patches. Code introduced while revising a fix is unneeded only slightly more often than code from the first fix, within model families: 0.26 vs.\ 0.19 for Claude, 0.24 vs.\ 0.21 for GPT, 0.50 vs.\ 0.41 for Gemini Flash, and 0.25 vs.\ 0.20 for all models except Gemini Flash. Revision steps are 1{,}512 of 15{,}190 labeled code units. On DeepSWE, \role{Fix} steps contribute 72\% of unneeded code and \role{Refine} steps 22\%, with unneeded rates of 0.24 and 0.27. The full role table is in Appendix~\ref{app:roletab}.

\subsection{RQ4 Details}
\label{app:rq4}

Of the 39 model pairs within 3\,pp resolve rate, 19 have non-overlapping 95\% CIs on \(\NP_{\mathrm{code}}\), with gaps up to 0.158. Sonnet~5 has \(\NP_{\mathrm{code}}=0.922\) [0.903, 0.942] and Gemini~3.7 Flash 0.789 [0.759, 0.820], which is 0.34 vs.\ 1.23 unneeded code units per task. For the all-unit \(\NP\), the Spearman correlation with resolve rate is \(+0.07\) (\(p=0.79\)).

\paragraph{Unit granularity.}
A patch with a single code unit has \(\NP_{\mathrm{code}}=1\) by definition, and such patches are common (62\% of Sonnet~5's, 39\% of Gemini~3.8 Flash's, 50\% of developer patches). On multi-unit patches only, or with units weighted by changed lines, the model ranking is unchanged (Spearman \(+0.97\) and \(+0.99\)) and resolve rate stays uncorrelated (\(-0.01\) and \(-0.07\)). The three Gemini Flash models stay significantly below developers when line-weighted. On multi-unit patches, developers average 0.729 (237 patches) and agents about 0.75 (\(\Ddev\approx+0.02\)); the Gemini Flash models stay significantly below (\(-0.064\), \(-0.047\), \(-0.059\)).

\paragraph{Task selection.}
Each model was analyzed on its own resolved tasks (482 distinct tasks; 114 shared by all 18 models). With task fixed effects and task-clustered standard errors, the three Gemini Flash models stay significantly below developers and ten models are significantly above. The fixed-effect ranking matches the raw one (Spearman \(+0.99\)) and is unrelated to resolve rate (\(-0.10\), \(p=0.70\)). Task difficulty has no significant effect (\(-0.008\), \(p=0.83\)), and log developer code units has a negative one (\(-0.133\), \(p<0.001\)). On the 99 tasks that all models resolve and that have complete grading, the Gemini Flash models are again significantly below developers.

\paragraph{Other harnesses.}
We pair each model's Copilot CLI patches with its final patches from other agents on tasks both resolved (\tabref{tab:generalization}). For Opus~5 on SWE-bench Verified, \(\NP_{\mathrm{code}}\) is 0.903 under the Copilot coding agent, 0.911 under VS Code agent mode, against 0.910 and 0.909 for Copilot CLI on the same tasks. Over three models in Claude Code (426 pairs), the changes are \(+0.004\) (Opus~5), \(+0.016\) (GPT-6 Astra), and \(-0.010\) (GPT-5.6 Luna). On DeepSWE, Codex CLI over six GPT models (179 pairs) gives \(+0.006\) [\(-0.01\), \(+0.03\)], and Claude Code over three Claude models (105 pairs) gives \(-0.007\) [\(-0.03\), \(+0.01\)].

\begin{table}[H]
\tablestyle{5pt}{1.15}
\begin{tabular}{llcccc}
\toprule
\textbf{Benchmark} & \textbf{Harness} & \textbf{Models} & \textbf{Resolve (\%)} & \(\NP_{\mathrm{code}}\) & \(r_s\) \\
\midrule
SWE-bench Verified & Copilot CLI & 18 & 64.0--89.4 & 0.887 & \(-0.10\) \\
DeepSWE & Copilot CLI & 17 & 21.2--68.1 & 0.797 & \(-0.59\) \\
SWE-bench Verified & Claude Code & 3 & -- & 0.926 (CLI 0.923) & -- \\
SWE-bench Verified & Copilot coding agent & 1 & -- & 0.903 (CLI 0.910) & -- \\
SWE-bench Verified & VS Code agent mode & 1 & -- & 0.911 (CLI 0.909) & -- \\
DeepSWE & Claude Code & 3 & -- & 0.828 (CLI 0.835) & -- \\
DeepSWE & Codex CLI & 6 & -- & 0.787 (CLI 0.794) & -- \\
\bottomrule
\end{tabular}
\caption{Necessity across benchmarks and harnesses. Resolve is the range across models, and \(r_s\) is the Spearman correlation between resolve rate and \(\NP_{\mathrm{code}}\) across models. The first two rows give the pooled mean over patches (6{,}728 and 932). The other rows use final patches only on tasks resolved under both harnesses, with the paired Copilot CLI (CLI) value in parentheses.}
\label{tab:generalization}
\end{table}

\paragraph{Across benchmarks.}
On DeepSWE the Spearman correlation between resolve rate and \(\NP_{\mathrm{code}}\) is \(-0.59\) (\(p=0.01\)), and \(-0.52\) (\(p=0.03\)) for the task-adjusted measure (each model's necessity minus the mean of all models on the same task, over the 92 tasks solved by at least three models). The raw model ranking agrees moderately with SWE-bench Verified (\(+0.59\), \(p=0.01\)), and the task-adjusted ranking slightly more (\(+0.63\), \(p=0.007\)). Gemini~3.7 and 3.8 Flash and GPT-6 Astra are significantly below the task mean, and Sonnet~5.5, Opus~4.8, Haiku~5.5, Kimi~K3, and Opus~5 are above it (Appendix~\ref{app:dswe}). On SWE-bench Pro, two models show no sign of over-editing relative to developers, but the Pro developer patches are far from test-minimal (Appendix~\ref{app:pro}).

\subsection{RQ5 Details}
\label{app:rq5}

\paragraph{First green run.}
Of 6{,}728 trajectories, 4{,}860 reach a first all-green run of the agent's own tests after its first source edit, and in 3{,}035 of them (62\%) the agent keeps editing. On the 3{,}031 graded patches, code units go from 7{,}419 to 7{,}269, and 2{,}948 (97.3\%) still pass the task tests. Of the 2{,}939 changed patches that still resolve, the post-green edits change source-code units in 208, touch only tests, documentation, or configuration in 1{,}472, and cancel out in 1{,}259.

\paragraph{Self-audit.}
We showed each resolved SWE-bench Verified patch, with the issue, to the same model (or the closest sibling) and asked it to justify every hunk and drop the rest. The audit covers 17 of 18 models (no Copilot-CLI auditor exists for Kimi~K3); where the original auditor model was unavailable a sibling model audited (Opus~5 by Claude Opus~5.5, Opus~4.8 by Claude Opus~4.6, GPT-5.6 Terra by GPT-6.1 Terra), and 4 over-length patches were skipped. The audit covers 6{,}345 patches and 13{,}729 code units; 1{,}540 patches had at least one drop. Against \(\Oracle^{+}\), 813 of the 1{,}238 dropped code hunks are unneeded (precision 0.657, recall 0.355, \(\kappa=0.389\); TP/FP/FN/TN = 813/425/1{,}478/11{,}013). Under \(\Oracle_{\mathrm{test}}\) precision is 0.785 (recall 0.288, \(\kappa=0.333\); 972/266/2{,}403/10{,}088), under \(\Oracle^{+}_{\mathrm{any}}\) 0.631 (recall 0.360; 781/457/1{,}391/11{,}100), and under \(\Oracle_{\mathrm{live}}\) 0.164 (recall 0.287; 203/1{,}035/504/11{,}987). Among \(\Oracle_{\mathrm{test}}\)-unneeded units, models drop 40.5\% of those that behave like the patch without them (unrequested changes), against 14.7\% of those that behave like the full patch (Fisher \(p=8\times10^{-46}\)) and 2.6\% of needed units. Pooled against the original \DRA{} grading (\(\Oracle_{\mathrm{test}}\)), 79\% of dropped code hunks are unneeded and the audit finds 29\% of the 3{,}407 unneeded code hunks (\(\kappa=0.33\)). Applying the drops lifts \(\NP_{\mathrm{code}}\) from 0.924 to 0.944 under \(\Oracle^{+}\), and 98.2\% of the regraded patches still pass \(\Oracle^{+}\) (the drops break 1.8\%; none were ungraded).

\paragraph{Pruning with \DRA{}.}
Of the 6{,}724 patches, 3{,}777 have no standalone test selector in the trajectory and 450 fail the baseline repeatability check (Appendix~\ref{app:cleanup-eligibility}). On the remaining 2{,}495, the cumulative leave-one-out update is \(S \leftarrow S \revert u\) whenever \(\Oracle_{\mathrm{agent}}(S \revert u)=1\), without emptying all code units. With \(\Oracle_{\mathrm{agent}}\), \(\NP_{\mathrm{code}}\) rises from 0.936 to 0.974, and 86.6\% of patches still pass the task tests (84.3\% without the empty-patch guard, at 37.3\% precision and 68.2\% recall). With \(\Oracle_{\mathrm{agent+P2P}}\), \(\NP_{\mathrm{code}}\) rises to 0.975, and 89.1\% still pass the task tests. The false drops are code units that implement behavior that neither the existing tests nor the agent's tests exercise.

\subsection{Validity Check Against Developer Patches}
\label{app:validity}

The outside-patch check uses 14{,}591 agent code units on 6{,}596 matched instances, of which 2{,}607 are outside-patch. Of the 3{,}753 units the metric flags, 43\% are outside-patch, and the rest touch developer-modified functions but add unrequested behavior or dead guards. Patch size cannot separate Sonnet~5 (0.86 lines per developer line, \(\NP_{\mathrm{code}}=0.922\)) from GPT-6.1 Sol (0.84, 0.887).

\section{Per-Model and Per-Benchmark Results}
\label{app:results-group}

\subsection{Cohort Funnel}
\label{app:funnel}

\tabref{tab:funnel} lists the nested denominators used in the paper, from the instances sampled to the cohorts behind each analysis.

\begin{table}[H]
\tablestyle{4pt}{1.12}
\begin{tabular}{rll}
\toprule
\textbf{Count} & \textbf{Cohort} & \textbf{Used for} \\
\midrule
6{,}728 & Resolved SWE-bench Verified patches analyzed (18 models) & Main results \\
6{,}345 & Resolved SWE-bench Verified patches in the self-audit (17 models; 13{,}729 code units) & RQ5 self-audit \\
5{,}482 & Patches that edit tests & RQ2 test analyses \\
2{,}945 & Patches with a standalone test selector, run in containers & \(\Oracle_{\mathrm{agent}}\) \\
2{,}495 & Repeatably passing \(\Oracle_{\mathrm{agent}}\) & \(\Oracle_{\mathrm{agent}}\) cleanup \\
3{,}035 & Patches that keep editing after first green (of 4{,}860 that reach it) & Post-green analysis \\
3{,}031 & Of these, graded & Post-green analysis \\
2{,}948 & Of these, pass the grader & Post-green analysis \\
\bottomrule
\end{tabular}
\caption{Nested cohorts. The main results analyze all 6{,}728 resolved patches; the cleanup experiments use the 6{,}724 resolved patches of 18 models, and the container cohorts (2{,}945, 2{,}495) are subsets of them.}
\label{tab:funnel}
\end{table}

\subsection{Per-Model Results}
\label{app:modelresults}

\tabref{tab:results} collects the per-model results on both benchmarks; \tabref{tab:results_otest} gives the confidence intervals.

\begin{table}[H]
\tablestyle{3.5pt}{1.1}
\begin{tabular}{lcccccccc}
\toprule
\textbf{Model} & \textbf{Resolve} & \textbf{Tasks} & \(\NP_{\mathrm{test}}\) & \(\NP_{\mathrm{code}}\) & \(\Ddev\) \textbf{vs.\ dev.} & \(\rho\) & \textbf{DeepSWE} \(\NP_{\mathrm{code}}\) & \textbf{Lines/dev.} \\
\midrule
Developer patch & n/a & 470 & 0.864 & 0.864 & -- & n/a & 0.785 & 1.00 \\
\midrule
Claude Opus 5.5 & 89.4 & 424 & 0.451 & 0.900 & \textbf{+.042} & 0.39 & 0.790 & 0.82 \\
Kimi K3 & 88.4 & 410 & 0.479 & 0.897 & +.024 & 0.36 & 0.822 & 0.86 \\
Claude Opus 5 & 87.2 & 420 & 0.657 & 0.898 & \textbf{+.031} & 0.25 & 0.808 & 0.94 \\
Claude Opus 4.8 & 85.0 & 404 & 0.657 & 0.912 & \textbf{+.045} & 0.34 & 0.866 & 0.91 \\
GPT-6 Astra & 84.4 & 400 & 0.279 & 0.889 & \textbf{+.025} & 0.24 & 0.749 & 0.85 \\
GPT-5.6 Sol & 83.6 & 388 & 0.422 & 0.924 & \textbf{+.042} & 0.02 & 0.787 & 0.84 \\
Gemini 3.8 Flash & 82.4 & 392 & 0.339 & 0.765 & \textbf{$-$.095} & 0.37 & 0.748 & 1.32 \\
Gemini 3.7 Flash & 81.4 & 392 & 0.348 & 0.789 & \textbf{$-$.074} & 0.39 & 0.730 & 1.36 \\
Claude Sonnet 5 & 80.6 & 382 & 0.634 & 0.922 & \textbf{+.052} & 0.34 & 0.814 & 0.86 \\
Claude Haiku 5.5 & 79.8 & 351 & 0.552 & 0.913 & \textbf{+.044} & 0.42 & 0.834 & 0.81 \\
GPT-6.1 Sol & 79.4 & 377 & 0.273 & 0.887 & +.024 & 0.18 & 0.767 & 0.84 \\
GPT-6 Sol & 78.6 & 363 & 0.333 & 0.886 & +.011 & 0.12 & 0.770 & 0.89 \\
GPT-5.6 Terra & 77.4 & 360 & 0.433 & 0.928 & \textbf{+.058} & 0.04 & 0.840 & 0.77 \\
Claude Sonnet 5.5 & 76.4 & 367 & 0.848 & 0.929 & \textbf{+.055} & 0.07 & 0.821 & 0.72 \\
GPT-5.6 Luna & 76.0 & 330 & 0.463 & 0.946 & \textbf{+.059} & 0.05 & 0.798 & 0.74 \\
Gemini 3.6 Flash & 73.2 & 349 & 0.374 & 0.791 & \textbf{$-$.076} & 0.38 & 0.801 & 1.36 \\
GPT-6 Luna & 71.6 & 312 & 0.431 & 0.921 & \textbf{+.048} & 0.08 & 0.805 & 0.79 \\
Claude Haiku 4.5 & 64.0 & 307 & 0.522 & 0.873 & $-$.003 & 0.38 & n/a & 1.15 \\
\midrule
All models & -- & 6{,}728 patches & 0.472 & 0.887 & -- & 0.257 & 0.797 & 0.86 \\
\bottomrule
\end{tabular}
\caption{Per-model results on GitHub Copilot CLI trajectories, SWE-bench Verified (all resolved instances; \(\Oracle_{\mathrm{test}}\)). \(\NP_{\mathrm{test}}\) is over all units; \(\NP_{\mathrm{code}}\) is over code units on each model's own instances. Tasks is the number of tasks analyzed per model. \(\Ddev\) is agent minus developer \(\NP_{\mathrm{code}}\) on matched instances; bold means the 95\% bootstrap CI excludes 0. \(\rho\) is the debug share of all residue. The DeepSWE developer baseline is 0.785 over 103 graded reference solutions; Haiku 4.5 solved one DeepSWE task. Lines/dev.\ is code lines changed by the agent divided by those changed by the developer on matched instances (ratio of sums; bottom row: median across models). Bottom-row \(\NP\) values are means over patches.}
\label{tab:results}
\end{table}

\subsection{Introducing Role of Unneeded Code}
\label{app:roletab}

\tabref{tab:roles} splits code units by the role of the step that introduced them.

\begin{table}[H]
\tablestyle{5pt}{1.1}
\begin{tabular}{lrrrrrr}
\toprule
& \multicolumn{3}{c}{SWE-bench Verified} & \multicolumn{3}{c}{DeepSWE} \\
\cmidrule(lr){2-4}\cmidrule(lr){5-7}
\textbf{Introducing role} & \textbf{Needed} & \textbf{Unneeded} & \textbf{Rate} & \textbf{Needed} & \textbf{Unneeded} & \textbf{Rate} \\
\midrule
Fix & 9{,}799 & 3{,}315 (83\%) & 0.25 & 7{,}869 & 2{,}547 (72\%) & 0.24 \\
Refine & 1{,}074 & 438 (11\%) & 0.29 & 2{,}146 & 796 (22\%) & 0.27 \\
Probe & 182 & 120 (3\%) & 0.40 & 494 & 78 (2\%) & 0.14 \\
Reproduce & 76 & 33 (1\%) & 0.30 & 122 & 52 (1\%) & 0.30 \\
Cleanup & 71 & 79 (2\%) & 0.53 & 477 & 81 (2\%) & 0.15 \\
\bottomrule
\end{tabular}
\caption{Code units by the role of their introducing step (labeled units only). Percentages are shares of unneeded code units; rate is unneeded divided by all units with that role.}
\label{tab:roles}
\end{table}

\subsection{DeepSWE Per-Model Results}
\label{app:dswe}

\begin{figure}[H]
\centering
\includegraphics[width=0.7\linewidth]{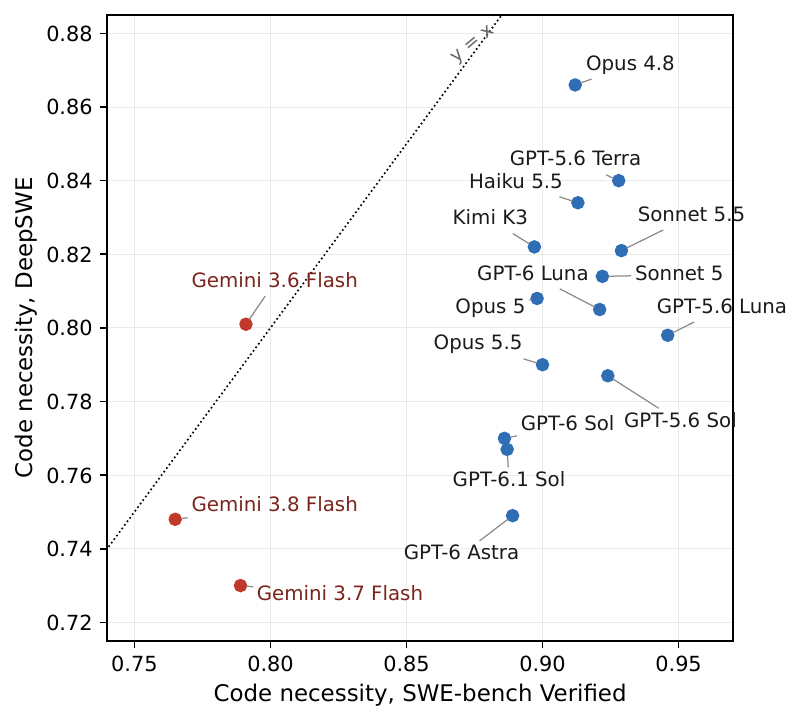}
\caption{Raw \(\NP_{\mathrm{code}}\) per model on SWE-bench Verified and DeepSWE (17 models with at least 10 graded DeepSWE tasks; red: Gemini Flash). Rank correlation is \(+0.59\) raw and \(+0.63\) after adjusting DeepSWE for task mix.}
\label{fig:crossdataset}
\end{figure}

\figref{fig:crossdataset} compares per-model \(\NP_{\mathrm{code}}\) across the two benchmarks, and \tabref{tab:dswe} lists the DeepSWE values.

\begin{table}[H]
\tablestyle{6pt}{1.12}
\begin{tabular}{lcccc}
\toprule
\textbf{Model} & \textbf{DeepSWE resolve} & \textbf{Tasks} & \textbf{Raw} \(\NP_{\mathrm{code}}\) & \textbf{Task-adjusted} [95\% CI] \\
\midrule
Claude Opus 4.8 & 49.6 & 47 & 0.866 & +0.045 [+0.024, +0.067] \\
Claude Sonnet 5.5 & 56.6 & 55 & 0.821 & +0.042 [+0.018, +0.066] \\
Claude Haiku 5.5 & 48.7 & 47 & 0.834 & +0.037 [+0.006, +0.067] \\
Kimi K3 & 66.4 & 64 & 0.822 & +0.032 [+0.016, +0.047] \\
Claude Opus 5 & 64.6 & 65 & 0.808 & +0.027 [+0.013, +0.042] \\
GPT-5.6 Luna & 21.2 & 23 & 0.798 & +0.023 [$-$0.020, +0.066] \\
GPT-5.6 Terra & 37.2 & 36 & 0.840 & +0.021 [$-$0.007, +0.046] \\
Claude Sonnet 5 & 41.6 & 43 & 0.814 & +0.017 [$-$0.013, +0.049] \\
GPT-6 Luna & 46.9 & 49 & 0.805 & +0.011 [$-$0.017, +0.037] \\
Gemini 3.6 Flash & 41.6 & 42 & 0.801 & +0.001 [$-$0.030, +0.034] \\
Claude Opus 5.5 & 67.3 & 68 & 0.790 & $-$0.011 [$-$0.037, +0.013] \\
GPT-5.6 Sol & 58.4 & 57 & 0.787 & $-$0.018 [$-$0.044, +0.007] \\
GPT-6 Sol & 60.2 & 58 & 0.770 & $-$0.020 [$-$0.049, +0.005] \\
GPT-6.1 Sol & 68.1 & 67 & 0.767 & $-$0.021 [$-$0.047, +0.005] \\
GPT-6 Astra & 67.3 & 63 & 0.749 & $-$0.033 [$-$0.062, $-$0.006] \\
Gemini 3.7 Flash & 68.1 & 63 & 0.730 & $-$0.047 [$-$0.071, $-$0.022] \\
Gemini 3.8 Flash & 65.5 & 64 & 0.748 & $-$0.049 [$-$0.075, $-$0.023] \\
\bottomrule
\end{tabular}
\caption{DeepSWE, Copilot CLI, sorted by task-adjusted \(\NP_{\mathrm{code}}\) (model necessity minus the mean of all models on the same task, over the 92 tasks solved by \(\ge\)3 models). Raw and adjusted values correlate at Spearman \(+0.92\). The developer baseline is 0.785 [0.749, 0.820] over 103 tasks. Claude Haiku~4.5 solved one task and is omitted.}
\label{tab:dswe}
\end{table}

\subsection{SWE-bench Pro (Supplementary)}
\label{app:pro}

We use complete Copilot CLI runs of two models on the SWE-bench Pro public set~\citep{deng2025swebenchpro}: GPT-5.6 Luna (399 of 731 tasks resolved) and GPT-5.3 Codex (316 of 731); later Pro runs failed at agent setup. We graded all resolved public-split tasks (398 Luna, 315 Codex), unitizing the final patches without replay (git-submodule pointer changes dropped); deltas use the 345/272 tasks where both the model and developer patches re-resolve under \DRA{} and grading completed (\tabref{tab:pro}).

\begin{table}[H]
\tablestyle{3pt}{1.15}
\begin{tabular}{lcccccc}
\toprule
\textbf{Model} & \textbf{Tasks} & \textbf{Model} & \textbf{Developer} & \(\Ddev\) [95\% CI] & \(\Ddev\), line-weighted & \textbf{Units/task} \\
\midrule
GPT-5.6 Luna & 345 & 0.609 & 0.499 & +0.111 [+0.08, +0.14] & +0.098 [+0.07, +0.12] & 9.0 vs.\ 12.4 \\
GPT-5.3 Codex & 272 & 0.597 & 0.481 & +0.116 [+0.09, +0.15] & +0.114 [+0.08, +0.15] & 9.0 vs.\ 12.5 \\
\bottomrule
\end{tabular}
\caption{SWE-bench Pro, Copilot CLI final patches vs.\ developer patches on the same tasks. Model and Developer columns give \(\NP_{\mathrm{code}}\); \(\Ddev\) is model minus developer (paired task bootstrap); units/task counts code units, model vs.\ developer.}
\label{tab:pro}
\end{table}

Both models have higher \(\NP_{\mathrm{code}}\) than the developer, also on multi-unit patches (\(+0.069\) and \(+0.079\)). Pro developer patches are large, and the hidden tests need less than half of their code (\(\NP_{\mathrm{code}}\) 0.48--0.50, against 0.86 on SWE-bench Verified), so the developer patch is a weak reference for minimality on Pro. The check shows only that these two models do not over-edit relative to developers on Pro; two models are too few to test the ranking question.

\subsection{Model Pool}
\label{app:models}

\tabref{tab:models} lists every model in the GitHub Copilot CLI evaluation pool with the number of resolved instances analyzed and the replay success rate.

\begin{table}[H]
\tablestyle{6pt}{1.15}
\begin{tabular}{lccc}
\toprule
\textbf{Model} & \textbf{Resolve} & \textbf{Resolved tasks analyzed} & \textbf{Replay success} \\
\midrule
Claude Opus 5.5 & 89.4 & 424 & 99\% (443/447) \\
Kimi K3 & 88.4 & 410 & 95\% (422/442) \\
Claude Opus 5 & 87.2 & 420 & 99\% (432/436) \\
Claude Opus 4.8 & 85.0 & 404 & 97\% (413/425) \\
GPT-6 Astra & 84.4 & 400 & 98\% (412/422) \\
GPT-5.6 Sol & 83.6 & 388 & 95\% (398/418) \\
Gemini 3.8 Flash & 82.4 & 392 & 99\% (406/412) \\
Gemini 3.7 Flash & 81.4 & 392 & 99\% (402/407) \\
Claude Sonnet 5 & 80.6 & 382 & 98\% (394/403) \\
Claude Haiku 5.5 & 79.8 & 351 & 99\% (367/369) \\
GPT-6.1 Sol & 79.4 & 377 & 98\% (389/397) \\
GPT-6 Sol & 78.6 & 363 & 95\% (375/393) \\
GPT-5.6 Terra & 77.4 & 360 & 95\% (366/387) \\
Claude Sonnet 5.5 & 76.4 & 367 & 99\% (378/382) \\
GPT-5.6 Luna & 76.0 & 330 & 89\% (338/380) \\
Gemini 3.6 Flash & 73.2 & 349 & 98\% (358/366) \\
GPT-6 Luna & 71.6 & 312 & 89\% (320/358) \\
Claude Haiku 4.5 & 64.0 & 307 & 98\% (312/320) \\
\bottomrule
\end{tabular}
\caption{Full Copilot CLI model pool on SWE-bench Verified. Replay success is the share of resolved instances whose trajectory replays exactly (6{,}925 of 7{,}164, 96.7\%); 197 control failures then leave 6{,}728 analyzed.}
\label{tab:models}
\end{table}

\section{Test Edits and Mutation Testing}
\label{app:mut-group}

\subsection{Test-Edit Utility and Mutation Testing Details}
\label{app:testmutation}

\tabref{tab:testutility} gives the full mutation results over the 271 valid instances (1{,}962 evaluated mutants).

\begin{table}[H]
\tablestyle{3.2pt}{1.04}
\begin{tabular}{lccc}
\toprule
\textbf{Metric} & \textbf{Needed (\(n=1\))} & \textbf{Unneeded (\(n=0\))} & \textbf{Full frame (271)} \\
\midrule
Evaluated code mutants & 1{,}725 & 237 & 1{,}962 \\
Killed by baseline tests in touched files (\(S_0\)) & -- & -- & 45.6\% \\
Killed by agent test units & 73.2\% & 31.2\% & 68.1\% \\
Killed by held-out task tests (\(\Oracle_{\mathrm{test}}\)) & 83.4\% & 31.2\% & 77.1\% \\
Killed by strengthened oracle (\(\Oracle^{+}\)) & -- & -- & 84.7\% \\
Net gain of agent tests over baseline file tests & -- & -- & 58.8\% \\
Marginal kills of agent tests beyond \(\Oracle_{\mathrm{test}}\) and \(S_0\) & 4.6\% & \textbf{9.3\%} & 5.2\% \\
Marginal kills of agent tests beyond \(\Oracle^{+}\) and \(S_0\) & -- & -- & 4.1\% \\
\bottomrule
\end{tabular}
\caption{Mutation testing of agent-patched code units on SWE-bench Verified over the 271 valid instances of the 387-instance eligible frame. The needed and unneeded columns are \(\Oracle_{\mathrm{test}}\)-needed and unneeded code units. Marginal kills are mutants killed by agent tests but by none of the listed suites, as a share of all evaluated mutants in the column (not of the mutants those suites miss). -- marks values not reported by the full-frame run.}
\label{tab:testutility}
\end{table}

\noindent Among the 1{,}341 test-editing patches with unneeded source code, 37.8\% of test units reference symbols unique to the unneeded code.

\paragraph{Four-arm \texttt{mutmut} protocol and mutant catalog.}
To evaluate the fault-detection value of agent-edited test units (\secref{sec:rq2}), we used \texttt{mutmut} (v3.8.1, built on \texttt{libcst} v1.9.0)~\citep{hovmoller2026mutmut,jia2011mutation,just2014mutants} to enumerate syntactic mutants restricted to the line ranges of agent-modified Python code units across all 6{,}728 resolved SWE-bench Verified patches (4{,}461 unique patched files; 198 files fail to parse). Because \texttt{mutmut}~3.8.1 skips most decorated functions by default, we subclassed \texttt{MutationVisitor} to strip \texttt{node.decorators} only while walking the function body, preserving mutations inside decorated methods without mutating decorator arguments. Across all patches, \texttt{mutmut} produces 96{,}790 mutants: 77{,}549 in \(\Oracle_{\mathrm{test}}\)-needed code units and 19{,}241 in \(\Oracle_{\mathrm{test}}\)-unneeded code units (\tabref{tab:mutmut_ops}).

For dynamic execution, we scaled to the 387-instance eligible frame of single-patch runs that edit both Python source and test files (3{,}525 patches, 11 repositories). Of these, 271 instances pass the validity gates G1--G4 (115 fail the G1 base guard, mostly environment differences in astropy, django and sphinx, and 1 fails G4): worktree tree-hash parity with \(S_T\), zero-mutation control pass on \(T\), \(\ge 1\) collected test in \(A_{\mathrm{scope}}\), and \(\ge 1\) evaluated non-stillborn mutant. Of 2{,}937 planned mutants, 1{,}962 were evaluated and non-stillborn (56 stillborn, 3 timeouts, 5 unresolved, 21 skipped by the run budget). Each mutant was tested under up to four arms:
\begin{enumerate}[leftmargin=1.5em, topsep=2pt, itemsep=2pt]
  \item \textbf{Held-out task tests (\(T\), \(\Oracle_{\mathrm{test}}\)):} the benchmark's \texttt{FAIL\_TO\_PASS} and \texttt{PASS\_TO\_PASS} tests run after resetting test files and applying the developer \texttt{test\_patch}.
  \item \textbf{Strengthened oracle (\(\Oracle^{+} = T \cup P\)):} \(\Oracle_{\mathrm{test}}\) combined with differential behavioral probes \(P\) (\secref{sec:behavior}), where \(P\) kills a mutant if the mutant alters any probe's return value or exception outcome relative to the unmutated patch \(S_T\). The probes are the GPT-6.1 Sol-judged PatchDiff probe scripts, executed in the task container on every mutant; 98 instances (485 of the 2{,}937 planned mutants) are probe-eligible.
  \item \textbf{Baseline tests in agent-touched files (\(B\)):} the unedited \(S_0\) versions of the test files modified by the agent.
  \item \textbf{Agent test files (\(A\)) and exact agent test units (\(A_{\mathrm{scope}}\)):} the agent-modified test files in \(S_T\), scoped to the specific test functions or classes added or edited by the agent's test units.
\end{enumerate}
Before running mutants on an instance, a zero-mutation control run recorded baseline passing and failing test IDs for each arm; a mutant is killed by an arm only if it causes a test that passed on the unmutated control to fail or error, and mutants that fail collection or time out at import are excluded as stillborn.

\begin{table}[H]
\tablestyle{3.0pt}{1.10}
\begin{tabular}{lrr}
\toprule
\textbf{\texttt{mutmut} operator class} & \textbf{6{,}728 catalog} & \textbf{Share} \\
\midrule
Call argument (\texttt{CALL})        & 41{,}890 & 43.3\% \\
String constant (\texttt{STR})       & 19{,}231 & 19.9\% \\
Operator swap (\texttt{OPR} / comp.) & 12{,}836 & 13.3\% \\
Assignment / null (\texttt{ASN})     & 10{,}321 & 10.7\% \\
Number constant (\texttt{NUM})       & 6{,}280 & 6.5\% \\
Unary / keyword / other              & 6{,}232 & 6.4\% \\
\midrule
All mutants                          & 96{,}790 & 100\% \\
\bottomrule
\end{tabular}
\caption{\texttt{mutmut}~3.8.1 operator distribution across all 6{,}728 resolved SWE-bench Verified patches. Unary / keyword / other combines unary negation (2{,}572), name/keyword swap (1{,}983), ternary swap (1{,}336), control-flow jump (257) and lambda body (84). Per-operator kill rates of the executed mutants, which use a different operator set, are in \tabref{tab:mutmut_kill}.}
\label{tab:mutmut_ops}
\end{table}

\paragraph{Full-frame results.}
\begin{sloppypar}
Across the 271 valid instances (1{,}962 evaluated mutants), \(B\) kills 45.6\% (95\% instance-cluster bootstrap CI \([41.1\%, 50.1\%]\)), \(A_{\mathrm{scope}}\) kills 68.1\% (\([63.7\%, 72.5\%]\); 79.7\% at whole-file level \(A\), \([76.2\%, 83.1\%]\)), and \(T\) kills 77.1\% (\([73.6\%, 80.5\%]\)). The net gain of \(A_{\mathrm{scope}}\) over \(B\) is 58.8\% (1{,}153/1{,}962, \([54.2\%, 63.3\%]\)). A marginal kill is a mutant that the reference suite misses and that at least one \(A_{\mathrm{scope}}\) test kills, excluding tests whose unedited version in \(B\) also kills it; because the exclusion is per test, a mutant can count even when a different baseline test also kills it. Marginal kills beyond \(T\) and \(B\) are 5.2\% (102/1{,}962, \([3.7\%, 6.9\%]\)), which is 22.7\% of the mutants \(T\) misses and occurs in 46 of 271 instances. Needed code is killed far more often than unneeded code under both \(A_{\mathrm{scope}}\) (73.2\% vs.\ 31.2\%) and \(T\) (83.4\% vs.\ 31.2\%); marginal kills beyond \(T\) are 4.6\% (80) on needed and 9.3\% (22) on unneeded code. The probes alone kill 12.2\% (239/1{,}962), so \(\Oracle^{+}\) kills 84.7\% (1{,}661/1{,}962, \([82.0\%, 87.2\%]\)); the stricter \(\Oracle^{+}_{\mathrm{strict}}\), which uses only probes from fully live units, kills 79.4\% (\([75.9\%, 82.6\%]\)). With the same exclusion of \(B\), agent tests still add kills beyond \(\Oracle^{+}\): 4.1\% (81/1{,}962, \([2.8\%, 5.6\%]\); 38 of 271 instances) and 4.6\% beyond \(\Oracle^{+}_{\mathrm{strict}}\) (91/1{,}962, \([3.2\%, 6.2\%]\); 43 instances). \tabref{tab:mutmut_kill} gives per-operator kill rates.
\end{sloppypar}

\begin{table}[H]
\tablestyle{3.0pt}{1.10}
\begin{tabular}{lrrr}
\toprule
\textbf{Operator} & \textbf{Mutants} & \textbf{\(T\) kill} & \textbf{\(\Oracle^{+}\) kill} \\
\midrule
Constant (CON) & 569 & 66.3\% & 77.2\% \\
Statement deletion (SDL) & 507 & 78.9\% & 86.4\% \\
Relational operator (ROR) & 195 & 81.5\% & 89.7\% \\
Condition (COND) & 244 & 88.5\% & 92.6\% \\
Arithmetic operator (AOR) & 111 & 83.8\% & 90.1\% \\
Return value (RET) & 143 & 78.3\% & 84.6\% \\
Logical connector (LCR) & 110 & 80.9\% & 85.5\% \\
Negation (NEG) & 83 & 80.7\% & 81.9\% \\
\midrule
All evaluated mutants & 1{,}962 & 77.1\% & 84.7\% \\
\bottomrule
\end{tabular}
\caption{Per-operator kill rates of the held-out task tests (\(T\)) and the strengthened oracle (\(\Oracle^{+}\)) over the 1{,}962 evaluated mutants. These operator classes are those of the executed mutants and differ from the static \texttt{mutmut} catalog classes in \tabref{tab:mutmut_ops}.}
\label{tab:mutmut_kill}
\end{table}

\paragraph{Static anatomy, timing, and per-model breakdown.}
Across all 6{,}728 SWE-bench Verified patches, 11{,}614 of 11{,}640 test units (99.8\%) have \(n(u)=0\) (9{,}617 in test files reset by \texttt{eval.sh} and 2{,}023 in other test files). On DeepSWE, which does not reset test files, 2{,}440 of 2{,}889 test units (84.5\%) still have \(n(u)=0\). Structurally, 8{,}702 of the 11{,}640 test units (74.8\%) define or modify a test function (detected by a function named \texttt{test\_*}), 1{,}179 (10.1\%) are import-only hunks, and 1{,}759 (15.1\%) edit helpers or fixtures; 72.0\% reside in or add an assertion-bearing test. Relative to the final needed code edit \(t^\star_{\mathrm{code}}\) across the 5{,}439 test-editing patches with step provenance, 2{,}827 (52.0\%) write all surviving test units strictly after \(t^\star_{\mathrm{code}}\), 1{,}704 (31.3\%) write them in the same step with none before, and 908 (16.7\%) write at least one before \(t^\star_{\mathrm{code}}\). Only 471 patches (8.7\%; 610 = 11.2\% raw) record a pre-fix red-to-green transition in bash. \tabref{tab:testmutation_models} gives the per-model breakdown: across the 18 models, writing tests before the fix correlates strongly with witnessing a red-to-green transition (Spearman \(r_s = 0.83, p < 10^{-4}\) raw; \(r_s = 0.80, p < 10^{-3}\) strict pre-fix).

\begin{table}[H]
\tablestyle{1.5pt}{1.10}
\resizebox{\linewidth}{!}{%
\begin{tabular}{lrrrrrrrrrr}
\toprule
\textbf{Model} & \textbf{Patches} & \textbf{Units} & \textbf{\texttt{test\_*}} & \textbf{Assert} & \textbf{Needed} & \textbf{Dev} & \textbf{Before Fix \%} & \textbf{Post-fix} & \textbf{R$\to$G (Pre/Raw)} & \textbf{Sites (N/U)} \\
\midrule
Claude Opus 5.5 & 396 & 661 & 76.1\% & 71.4\% & 70.3\% & 61.4\% & 7.9\% & 68.3\% & 0.5\% / 1.8\% & 2{,}277 / 440 \\
Kimi K3 & 322 & 711 & 73.7\% & 70.7\% & 63.7\% & 58.9\% & 3.8\% & 96.2\% & 2.2\% / 4.1\% & 2{,}420 / 566 \\
Claude Opus 5 & 181 & 386 & 74.6\% & 73.6\% & 68.1\% & 59.6\% & 11.1\% & 81.1\% & 1.1\% / 2.8\% & 2{,}629 / 772 \\
Claude Opus 4.8 & 203 & 387 & 76.5\% & 74.4\% & 70.5\% & 64.9\% & 6.9\% & 92.1\% & 2.5\% / 4.4\% & 2{,}610 / 619 \\
GPT-6 Astra & 400 & 995 & 73.3\% & 71.1\% & 65.8\% & 59.3\% & 62.6\% & 19.6\% & 38.2\% / 44.7\% & 1{,}765 / 542 \\
GPT-5.6 Sol & 387 & 770 & 72.5\% & 69.9\% & 66.9\% & 58.7\% & 7.5\% & 19.7\% & 3.6\% / 5.2\% & 1{,}790 / 300 \\
Gemini 3.8 Flash & 391 & 1050 & 75.1\% & 72.5\% & 65.7\% & 60.1\% & 11.5\% & 92.4\% & 5.5\% / 8.7\% & 2{,}665 / 1{,}609 \\
Gemini 3.7 Flash & 391 & 1056 & 76.1\% & 75.5\% & 66.7\% & 59.0\% & 6.2\% & 95.6\% & 2.3\% / 4.4\% & 2{,}628 / 1{,}640 \\
Claude Sonnet 5 & 195 & 389 & 76.3\% & 73.8\% & 71.2\% & 65.6\% & 6.2\% & 93.8\% & 2.1\% / 3.6\% & 1{,}803 / 422 \\
Claude Haiku 5.5 & 270 & 463 & 78.6\% & 73.9\% & 70.0\% & 60.0\% & 10.9\% & 61.8\% & 1.9\% / 4.9\% & 2{,}311 / 373 \\
GPT-6.1 Sol & 377 & 956 & 70.5\% & 68.0\% & 66.9\% & 59.5\% & 48.7\% & 29.4\% & 29.1\% / 33.4\% & 1{,}731 / 457 \\
GPT-6 Sol & 363 & 805 & 73.8\% & 70.7\% & 64.8\% & 56.4\% & 20.7\% & 46.0\% & 11.8\% / 16.5\% & 1{,}877 / 511 \\
GPT-5.6 Terra & 358 & 711 & 70.3\% & 68.5\% & 66.4\% & 61.0\% & 10.9\% & 24.1\% & 6.2\% / 7.0\% & 2{,}262 / 262 \\
Claude Sonnet 5.5 & 45 & 101 & 87.1\% & 88.1\% & 69.3\% & 54.5\% & 2.3\% & 65.1\% & 0.0\% / 2.3\% & 2{,}074 / 460 \\
GPT-5.6 Luna & 319 & 549 & 76.0\% & 71.2\% & 68.5\% & 57.2\% & 11.6\% & 26.1\% & 6.6\% / 7.5\% & 1{,}420 / 225 \\
Gemini 3.6 Flash & 348 & 674 & 79.7\% & 75.1\% & 66.2\% & 58.5\% & 6.7\% & 94.2\% & 3.2\% / 3.8\% & 2{,}507 / 1{,}463 \\
GPT-6 Luna & 310 & 552 & 76.3\% & 72.5\% & 65.9\% & 57.8\% & 15.8\% & 31.3\% & 7.4\% / 9.0\% & 1{,}425 / 224 \\
Claude Haiku 4.5 & 226 & 424 & 75.2\% & 72.6\% & 73.3\% & 66.3\% & 16.8\% & 88.9\% & 9.3\% / 14.2\% & 1{,}671 / 458 \\
\midrule
All models & 5{,}482 & 11{,}640 & 74.8\% & 72.0\% & 67.2\% & 59.8\% & 16.7\% & 59.3\% & 8.7\% / 11.2\% & 37{,}865 / 11{,}343 \\
\bottomrule
\end{tabular}}
\caption{Per-model test-edit anatomy, AST symbol coupling, trajectory timing, and first-order syntactic AST mutation sites on SWE-bench Verified. Columns 4--7 are percentages of test units defining/modifying test functions (detected by a function named \texttt{test\_*}), containing assertions, referencing needed source symbols, and referencing developer-patch symbols; Columns 8--10 are percentages of labeled test-editing patches with any test unit written before the fix (\(t^\star_{\mathrm{test}} < t^\star_{\mathrm{code}}\)), with any test unit written post-fix (\(t^\star_{\mathrm{test}} > t^\star_{\mathrm{code}}\)), and witnessing a bash red-to-green transition (strict pre-fix / raw trajectory); Column 11 lists first-order syntactic AST mutation sites in needed (N) and unneeded (U) code units across all code units of each model.}
\label{tab:testmutation_models}
\end{table}

\section{Agent-Test Oracle and Removal Bounds}
\label{app:cleanup-group}

\subsection{Repeatably Passing Agent-Test Oracle (\texorpdfstring{\(\Oracle_{\mathrm{agent}}\)}{O\_agent}) and Non-Passing Cases}
\label{app:cleanup-eligibility}

\paragraph{Definition of repeatably passing (\(\Oracle_{\mathrm{agent}}(S_T)=1\)).}
A resolved patch \(S_T\) has a \emph{repeatably passing} agent-test oracle (\(\Oracle_{\mathrm{agent}}(S_T)=1\)) if the standalone test command extracted from the agent's trajectory still exists in \(S_T\) and passes twice consecutively with a non-zero, identical passing test count when re-executed against the unpruned final patch \(S_T\) before reverting any units. Because post-generation cleanup tests each candidate revert by checking whether \(\Oracle_{\mathrm{agent}}(S \revert u)=1\) (requiring the passing test count to match baseline), verifying \(\Oracle_{\mathrm{agent}}(S_T)=1\) on the unpruned patch is a prerequisite: if \(\Oracle_{\mathrm{agent}}(S_T)=0\), every candidate revert would fail the oracle check and no unit could be pruned.

\paragraph{Label-free eligibility funnel across the 6{,}724 resolved patches.}
Because the cleanup procedure never reverts the last remaining code unit (\(|S \cap \Units_{\mathrm{code}}| \ge 1\)), all 3{,}781 single-code-unit patches (\(|S_T \cap \Units_{\mathrm{code}}|=1\)) are deterministic no-ops from patch structure alone, without consulting \(\Oracle^{+}\) or \(\Oracle_{\mathrm{test}}\). \tabref{tab:cleanup-funnel} breaks down all 6{,}724 resolved patches (3{,}781 single-code-unit and 2{,}943 multi-code-unit) through trajectory test extraction and in-container baseline repeatability testing (executing C/Cython extension repositories directly in the pre-built \texttt{/testbed} environment so zero tasks fail environment imports):
\begin{enumerate}[leftmargin=1.5em, topsep=2pt, itemsep=2pt]
  \item \textbf{No standalone extracted test selector (3{,}777 of 6{,}724 patches).} The trajectory parser extracts new standalone test selectors without reading \(\Oracle^{+}\) or \(\Oracle_{\mathrm{test}}\); these patches either submit no test-file edits in \(S_T\) (for example, the agent ran a temporary reproduction script or an inline \texttt{python -c} command and deleted it) or modify existing test files without a standalone \texttt{pytest} or \texttt{runtests.py} node selector after the final edit.
  \item \textbf{In-container baseline check fails, \(\Oracle_{\mathrm{agent}}(S_T)=0\) (450 patches).} Of the 2{,}945 container-tested patches where \(\Oracle_{\mathrm{agent}}(S_T)\) is executed, 2{,}495 pass repeatably (\(\Oracle_{\mathrm{agent}}(S_T)=1\): 1{,}468 single-code-unit no-ops and 1{,}027 multi-code-unit patches) and 450 fail and are left untouched. The most common causes are runner mismatches in standalone replay (for example SymPy patches whose probe invokes \texttt{python -m pytest} where the agent ran \texttt{bin/test}), abstract mixin test classes in Django, strict pytest mark collection errors in Astropy, and extracted commands that relied on uncommitted scratch settings.
\end{enumerate}

\begin{table}[H]
\tablestyle{4pt}{1.12}
\begin{tabular}{lrrr}
\toprule
\textbf{Stage / Outcome} & \textbf{All patches} & \(|\Units_{\mathrm{code}}|=1\) \textbf{(no-op)} & \(|\Units_{\mathrm{code}}| \ge 2\) \\
\midrule
All resolved patches analyzed & 6{,}724 & 3{,}781 & 2{,}943 \\
\quad No standalone extracted test selector in trajectory & 3{,}777 & -- & -- \\
Container-verified patches tested with \(\Oracle_{\mathrm{agent}}(S_T)\) & 2{,}945 & -- & -- \\
\quad Baseline fails (\(\Oracle_{\mathrm{agent}}(S_T)=0\); left untouched) & 450 & -- & -- \\
\quad \textbf{Repeatably passing (\(\Oracle_{\mathrm{agent}}(S_T)=1\); evaluated)} & \textbf{2{,}495} & \textbf{1{,}468} & \textbf{1{,}027} \\
\bottomrule
\end{tabular}
\caption{Label-free eligibility breakdown from all 6{,}724 resolved SWE-bench Verified patches to the 2{,}495 container-evaluated patches with \(\Oracle_{\mathrm{agent}}(S_T)=1\). The single- and multi-code-unit split (\(|\Units_{\mathrm{code}}|=1\) vs.\ \(|\Units_{\mathrm{code}}| \ge 2\)) is shown where available, and is computed without consulting \(\Oracle^{+}\) or \(\Oracle_{\mathrm{test}}\).}
\label{tab:cleanup-funnel}
\end{table}

\paragraph{Pre-existing tests alone and full-suite reference bounds under \(\Oracle^{+}\).}
Because pre-existing regression tests (\(\Oracle_{\mathrm{P2P}}\), the PASS\_TO\_PASS suite) do not check the new issue fix or untested developer-aligned behavior, \(\Oracle_{\mathrm{P2P}}\) alone drops 2{,}540 \(\Oracle^{+}\)-needed units across all 6{,}724 patches (and 951 across the 2{,}495 container cohort; \tabref{tab:cleanup-bounds}). Guarded \(\Oracle_{\mathrm{agent}}\) in an oracle-guided simulation across all 6{,}724 patches reaches 57.5\% precision and 69.9\% recall, with 87.4\% of patches still passing \(\Oracle^{+}\), compared with 51.5\% precision and 85.4\% retention for the patch's own tests alone; of the 2{,}717 unneeded units, 1{,}899 (69.9\%) are pruned, 617 (22.7\%) are locked in by agent tests, and 201 (7.4\%) are kept by the empty-patch guard. Intersecting the in-container \(\Oracle_{\mathrm{agent+P2P}}\) drops with the self-audit on the 2{,}344 patches where both exist raises precision to 67.7\% at 98.3\% retention. Pruning with the strengthened oracle \(\Oracle^{+}\) keeps 99.8\% of patches passing (6{,}710 of 6{,}724) when all 2{,}717 unneeded units are removed simultaneously, and 100\% (99.5\% recall, 2{,}703 of 2{,}717 units) under guarded cumulative leave-one-out removal.

\begin{table}[H]
\tablestyle{3pt}{1.12}
\begin{tabular}{lccccccc}
\toprule
\textbf{Method} & \textbf{Patches} & \textbf{Dropped} & \textbf{TP / FP} & \textbf{Prec.} & \textbf{Recall} & \textbf{Post} \(\NP_{\mathrm{code}}\) & \textbf{Passes} \(\Oracle^{+}\) \\
\midrule
\DRA{} + \(\Oracle_{\mathrm{P2P}}\) (all) & 6{,}724 & 4{,}047 & 1{,}507 / 2{,}540 & 37.2\% & 55.5\% & 0.953 & 80.7\% \\
\DRA{} + \(\Oracle_{\mathrm{P2P}}\) (2{,}495 cohort) & 2{,}495 & 1{,}220 & 269 / 951 & 22.0\% & 45.3\% & 0.960 & 80.0\% \\
\DRA{} + \(\Oracle_{\mathrm{patch\text{-}test}}\) (sim.) & 6{,}724 & 3{,}449 & 1{,}775 / 1{,}674 & 51.5\% & 65.3\% & 0.962 & 85.4\% \\
\DRA{} + \(\Oracle_{\mathrm{agent}}\) (sim.) & 6{,}724 & 3{,}305 & 1{,}899 / 1{,}406 & 57.5\% & 69.9\% & 0.969 & 87.4\% \\
\DRA{} + \(\Oracle_{\mathrm{agent+P2P}}\) \(\cap\) audit & 2{,}344 & 229 & 155 / 74 & 67.7\% & 27.1\% & -- & 98.3\% \\
\DRA{} + \(\Oracle^{+}\) (cumulative) & 6{,}724 & 2{,}703 & 2{,}703 / 0 & 100.0\% & 99.5\% & 0.998 & 100.0\% \\
\bottomrule
\end{tabular}
\caption{Reference bounds for post-generation removal on SWE-bench Verified evaluated under \(\Oracle^{+}\) using pre-existing regression tests alone (\(\Oracle_{\mathrm{P2P}}\)), oracle-guided simulations over all 6{,}724 resolved patches (15{,}191 code units, 2{,}717 unneeded), and the full strengthened oracle (\(\Oracle^{+}\)). Post-reduction \(\NP_{\mathrm{code}}\) is the task mean. The \(\cap\) audit row is on the 2{,}344 patches that have both an in-container run and a self-audit.}
\label{tab:cleanup-bounds}
\end{table}

\section{Prompts and Protocols}
\label{app:prompts-group}

\subsection{Intervention Prompts (Not Run)}
\label{app:prompts}

These prompts were written for the intervention study that we did not run. The offline self-audit of \secref{sec:mechanism} adapts I2 to final patches, without tools.

\noindent\begin{minipage}{\linewidth}
\paragraph{I1: Revert-on-reject.} Appended to the system prompt:
\begin{lstlisting}
When you edit code to test a hypothesis and the hypothesis turns out
to be wrong, revert that edit before trying the next idea. Use
`checkpoint save <name>` before exploratory edits and
`checkpoint restore <name>` to undo them. Reproduction scripts and
debugging prints must not remain in the final working tree.
\end{lstlisting}
\end{minipage}

\noindent\begin{minipage}{\linewidth}
\paragraph{I2: Pre-submit diff audit.} Sent before submission:
\begin{lstlisting}
Before submitting, here is your complete diff against the original
repository:
{full_diff}
For each hunk, state in one line why it is required to resolve the
issue. Remove every hunk you cannot justify, then submit.
\end{lstlisting}
\end{minipage}

\section{Extended Related Work}
\label{app:related}

\paragraph{Measuring over-editing.} Recent work measures how much of an agent's patch is unneeded by comparing it with a reference patch, using edit distance or patch-level diffs~\citep{zhu2026overediting,luo2026recap}. These measures treat every difference from the reference as unneeded. Our measure grades each unit with an oracle, so a different but valid edit is not counted against the agent.

\paragraph{Training for minimal edits.} Synthetic task generation~\citep{yang2025swesmith} and reinforcement learning with verifiable rewards~\citep{shao2024deepseekmath} could be shaped to reward smaller patches. Our results suggest what to target: unneeded code is mostly part of the fix, not debugging scaffolding.

\paragraph{Agent harnesses and trajectories.} Open harnesses and their published trajectories~\citep{minisweagent,wang2025openhands,swebenchexperiments} let us compare patches across agents. We replay trajectories to attribute each line to the step that wrote it, which final-patch comparisons cannot do.

\paragraph{Delta debugging and repair minimality.} Delta debugging reduces a failing input to a minimal one~\citep{zeller2002dd}, and repair work favors small changes that keep the program correct~\citep{mechtaev2015directfix}, though test-passing repairs can still overfit~\citep{smith2015cure}. Our oracle-guided unit removal follows the same idea for agent patches.

\paragraph{Tests and mutation testing.} \begin{sloppypar}Property-based testing~\citep{maciver2019hypothesis}, LLM-written differential tests~\citep{hu2026difftestgen}, and mutation testing~\citep{jia2011mutation,just2014mutants,hovmoller2026mutmut} measure whether tests detect changes in behavior. We use them to strengthen the oracle and to ask whether the agent's own tests guard its code.\end{sloppypar}